\documentclass[12pt,preprint]{aastex}
\usepackage{emulateapj5}
\usepackage{apjfonts}
\usepackage{epsfig}
\usepackage{natbib}
\newcommand{\alp}{\ensuremath{\alpha}}

\newcommand{\feh}{[Fe/H]}
\newcommand{\feave}{\ensuremath{\langle {\rm Fe} \rangle}}

\def\gtrsim{\mathrel{\hbox{\rlap{\hbox{\lower4pt\hbox{$\sim$}}}\hbox{\raise2pt\hbox{$>$}}}}}

\newcommand{\hii}{\ion{H}{2}}
\newcommand{\hbeta}{H\ensuremath{\beta}}

\newcommand{\kms}{km~s\ensuremath{^{-1}}}

\newcommand{\mgb}{Mg$b$}

\newcommand{\msun}{\ensuremath{M_{\odot}}}

\newcommand{\oiii}{[\ion{O}{3}]}

\newcommand{\sers}{S{\'e}rsic}

\newcommand{\sigmastar}{\ensuremath{\sigma_{\ast}}}

\def\lax{{$\mathrel{\hbox{\rlap{\hbox{\lower4pt\hbox{$\sim$}}}\hbox{$<$}}}$}}
\def\gax{{$\mathrel{\hbox{\rlap{\hbox{\lower4pt\hbox{$\sim$}}}\hbox{$>$}}}$}}

\slugcomment{To be published in {\it The Astrophysical Journal}.}
\shorttitle{{\it Elliptical Outskirts II}}
\shortauthors{GREENE, ET AL.}

\begin{document}

\title{The Stellar Halos of Massive Elliptical Galaxies II: 
Detailed Abundance Ratios at Large Radius}

\author{Jenny E. Greene\altaffilmark{1,3}, Jeremy D. Murphy\altaffilmark{1,4},
Genevieve J. Graves\altaffilmark{1}, James E. Gunn\altaffilmark{1}, 
Sudhir Raskutti\altaffilmark{1}, 
Julia M. Comerford\altaffilmark{2,4}, Karl Gebhardt\altaffilmark{2}}

\altaffiltext{1}{Department of Astrophysics, Princeton University}
\altaffiltext{2}{Department of Astronomy, UT Austin, 1 University Station C1400, 
Austin, TX 71712}
\altaffiltext{3}{Alfred P. Sloan Fellow}
\altaffiltext{4}{National Science Foundation Fellow}

\begin{abstract}
    We study the radial dependence in stellar populations of 33 nearby
  early-type galaxies with central stellar velocity dispersions
  \sigmastar$\gtrsim 150$~\kms. We measure stellar population 
  properties in composite spectra, and use ratios 
  of these composites to highlight the largest spectral changes as a function 
  of radius. Based on stellar population modeling,
  the typical star at $2 R_e$ is old ($\sim 10$ Gyr), relatively metal
  poor (\feh$\approx -0.5$), and \alp-enhanced ([Mg/Fe]$\approx 0.3$).
  The stars were made rapidly at $z \approx 1.5-2$ in shallow
  potential wells. Declining radial gradients in [C/Fe], which follow
  [Fe/H], also arise from rapid star formation timescales due to
  declining carbon yields from low-metallicity massive stars. In contrast, [N/Fe] 
  remains high at large radius. Stars at
  large radius have different abundance ratio patterns from stars in
  the {\it center} of any present-day galaxy, but are similar to
  Milky Way thick disk stars.  Our observations are thus
  consistent with a picture in which the stellar outskirts are built
  up through minor mergers with disky galaxies whose star formation is
  truncated early ($z \approx 1.5-2$).
\end{abstract}

\section{Introduction}

Galaxies exhibit strong correlations between their mass and their
metallicity.  It is thought that mass-metallicity correlations arise
from star-formation--driven winds preferentially removing metals from
low-mass galaxies
\citep[e.g.,][]{larson1974,dekelwoo2003,tremontietal2004}.  In the
case of elliptical galaxies, the mass-metallicity relation is
manifested most strongly in the \mgb-\sigmastar\ relation
\citep{dressleretal1987,benderetal1993}.  If indeed elliptical
galaxies grow through merging, as expected in a hierarchical universe,
then naively we have an accounting problem -- how to make massive
metal-rich galaxies through the addition of smaller metal-poor units
\citep{faberetal2007,naabostriker2009}.

The situation is more nuanced since (a) mass-metallicity relations
evolve with cosmic time, meaning that all galaxies had lower
metallicities in the past
\citep[e.g.,][]{erbetal2006,mannuccietal2010} and (b) galaxies have
known radial stellar population gradients.  In general, the
\mgb-\sigmastar\ relation is measured in the high surface-brightness
galaxy center, where the metallicity is highest.  A full census of
the metallicity and abundance ratio content of elliptical galaxies
requires spatially resolved observations.  The recent discovery of
dramatic (factor of 2-4) size growth in elliptical galaxies from $z
\approx 2$ to the present
\citep[e.g.,][]{daddietal2005,trujilloetal2006,vandokkumetal2008,vanderweletal2008,
  cimattietal2008,damjanovetal2009,williamsetal2010,cassataetal2010}
is now supported by an increasing number of dynamical studies
\citep{cappellarietal2009,vandesandeetal2011}.  It seems that
much of the late-time growth of elliptical galaxies has occurred in
their outer parts \citep[e.g.,][]{naabetal2009, vandokkumetal2010}.
If elliptical galaxies are formed in two phases, with an early
rapid gas-rich phase making the central compact galaxy observed at
high redshift, and a late-time accretion phase building up the outer
parts \citep[e.g.,][]{oseretal2010,hilzetal2012,hilzetal2013}, we might hope to see
the imprint of these two phases in the stellar population gradients.

In principle, the stellar outskirts carry important information about
the late-time assembly history of elliptical galaxies.  In practice,
observations of the stellar populations in elliptical galaxy outskirts
are challenging, since their surface brightnesses drop steeply with
radius.  Despite more than thirty years of effort, most observations
of stellar population gradients do not extend much beyond the
half-light radius
\citep{spinradtaylor1971,faberetal1977,gorgasetal1990,
  fisheretal1995,kobayashiarimoto1999,ogandoetal2005, broughetal2007,
  baesetal2007,annibalietal2007,sanchez-blazquezetal2007,rawleetal2008,
  kuntschneretal2010,coccatoetal2010,coccatoetal2011}.  
While integral-field spectrographs have
brought a golden age in the study of spatially resolved galaxy
properties \citep{emsellemetal2004,
  sarzietal2006,cappellarietal2006,cappellarietal2012}, there are
still few observations that extend beyond the half-light radius in
integrated light
\citep{carolloetal1993,carollodanziger1994,mehlertetal2003,
  kelsonetal2006,weijmansetal2009,spolaoretal2010,
  puetal2010,puhan2011}. To our knowledge, there are fewer than thirty
integrated-light observations in total (using different instruments
and techniques) that reach beyond $R_e$ in massive elliptical galaxies
in the literature.  There are also a handful of studies that reach
into elliptical galaxy halos using resolved stellar population studies
\citep[e.g.,][]{kaliraietal2006,harrisetal1999,rejkubaetal2005,harrisetal2007,crnojevicetal2013}.
We thus present the largest and most homogeneous spectroscopic sample
to date of observations $\gtrsim 2 R_e$.

Specifically, building on the preliminary study of \citet[][ Paper I,
hereafter]{greeneetal2012}, we use integral-field observations taken with the
Mitchell Spectrograph at McDonald Observatory. We measure robust
stellar population gradients out to 2.5$R_e$ in massive local
elliptical galaxies. With a $107\arcsec \times 107$~\arcsec\ field-of-view,
the Mitchell spectrograph is uniquely suited to explore massive galaxy
halos. Our sample comprises 33 galaxies, eight of which were already presented 
in Paper I.

We present the sample in \S \ref{sec:Sample}, the observations and
data reduction in \S \ref{sec:Obs}, our analysis in \S
\ref{sec:Analysis}, and the radial variations in stellar populations
in \S \ref{sec:Radial}.  We discuss our findings in the context of the
hierarchical assembly of massive galaxies in \S \ref{sec:Discussion},
and summarize in \S \ref{sec:Summary}.

\begin{figure*}
\hskip 15mm
\includegraphics[scale=.60,angle=90]{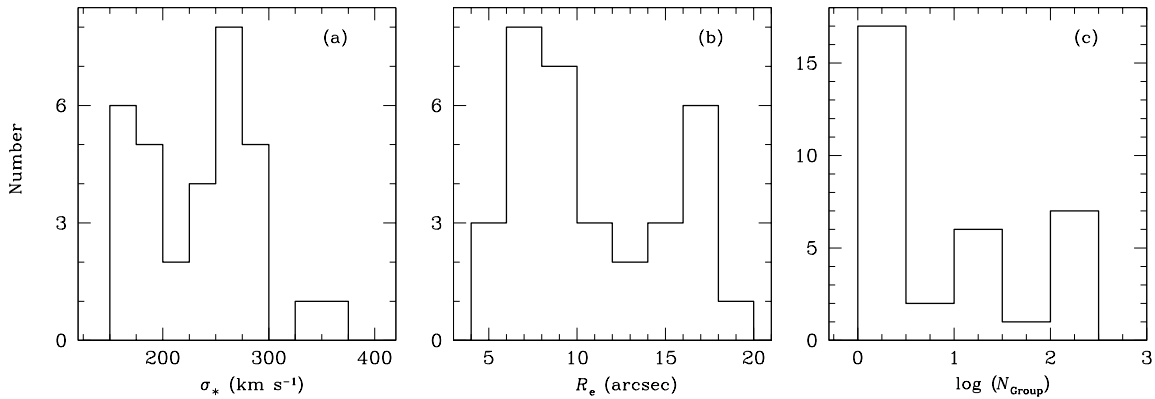}
\caption{{\bf a}:
Distribution of central \sigmastar\ (\kms) as measured by the SDSS for the entire 
sample presented in this paper.
{\bf b}:
Distribution in effective radius (\arcsec).  At the typical distance ($\sim 70$ Mpc) of our 
sample, 8\arcsec$\approx 2.5$ kpc. 
{\bf c}:
Distribution of the log of the number of group members for each galaxy 
\citep{yangetal2007,zhuetal2010,wetzeletal2012}.  
Note that this is a global measurement of environment.
\label{fig:histprop}
}
\end{figure*}

\begin{figure*}
\vbox{ 
\hskip 23mm
\psfig{file=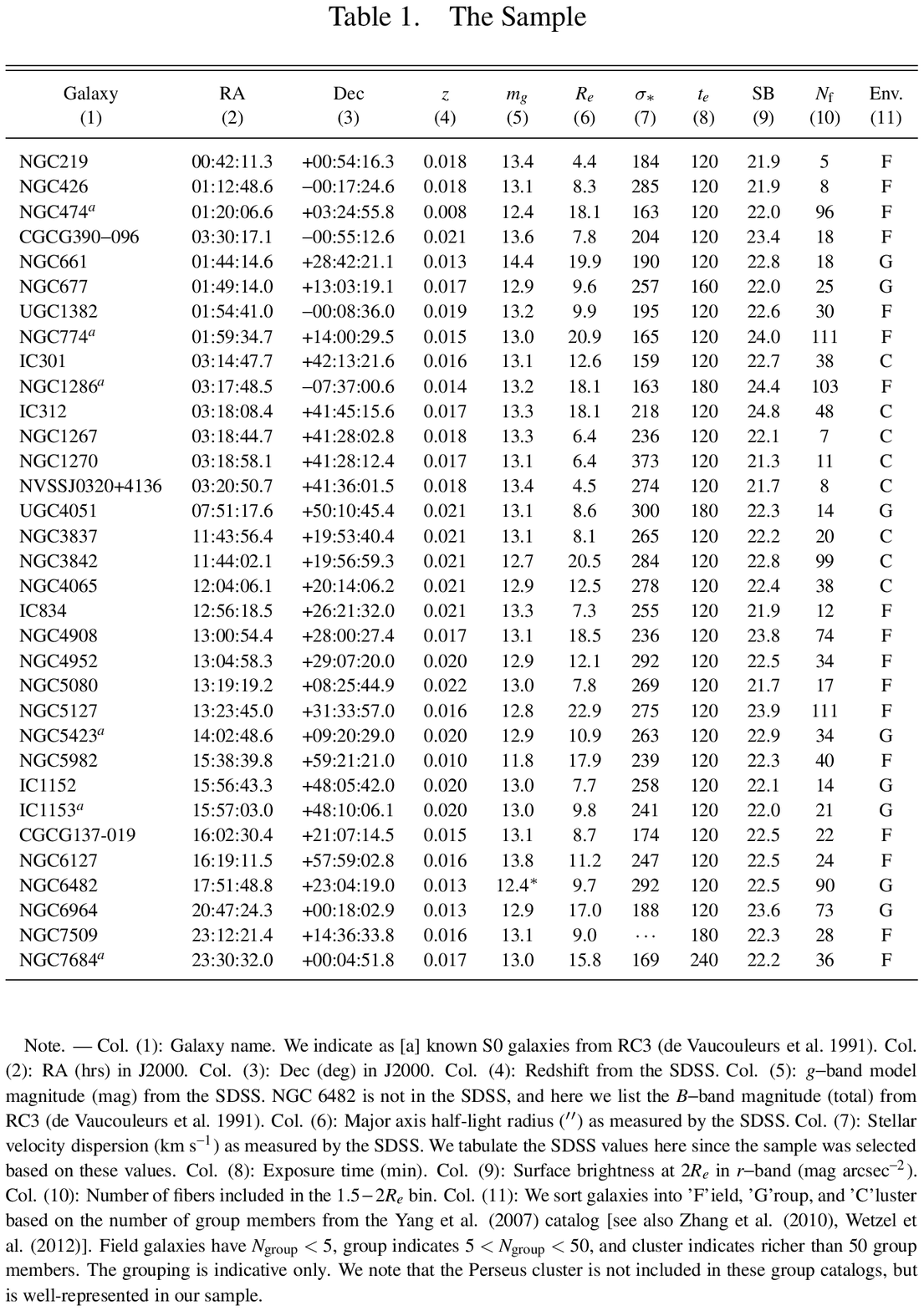,width=0.8\textwidth,keepaspectratio=true,angle=0}
}
\label{tab:obs}
\end{figure*}
\vskip 4mm

\section{Sample}
\label{sec:Sample}

We select our sample of massive early-type galaxies from the Sloan
Digital Sky Survey \citep[SDSS;][]{yorketal2000}.  The spectral
resolution of the Mitchell Spectrograph is $\sigma_{\rm inst} \approx
150$~\kms\ at $4700$~\AA; we select galaxies with dispersion measurements from the
SDSS that are greater than this value. Individual fibers are 4\arcsec\
in diameter, and so we aim for galaxies with effective radii at least
twice as large.  Galaxies with distances of 40-95 Mpc are large enough
to be well-resolved but small enough to fit into one pointing.  We use
a color selection of $u-r>2.2$ \citep[][]{stratevaetal2001}, which
preferentially selects early-type galaxies, and then remove the few
edge-on disk galaxies by hand.  There is no clean way to remove S0s;
we keep them in the sample and eventually hope to use kinematic
information to cleanly identify them. For now, we note the
photometrically classified S0 galaxies from the RC3
\citep{devaucouleursetal1991} in Table \ref{tab:obs}, where we list the full 
galaxy sample.  Finally, we use
the group catalogs of \citet{yangetal2007}, \citet{zhuetal2010}, 
and \citet{wetzeletal2012} to
get a global estimate of the galaxy environment.  We show the full
distribution of \sigmastar, half-light radius, and group membership for each galaxy in
the sample in Figure \ref{fig:histprop}.

\section{Observations and Data Reduction}
\label{sec:Obs}

The new observations of 25 galaxies presented in this paper were
observed over three runs in January 17-20 2012, May 20-24 2012, and
October 15-18 2012.  We include here also the eight galaxies presented
in \citet{greeneetal2012}, for a total sample of 33 galaxies. The
observations were made with the George and Cynthia Mitchell
Spectrograph \citep[the Mitchell Spectrograph, formerly
VIRUS-P;][]{hilletal2008a} on the 2.7m Harlan J. Smith telescope at
McDonald Observatory. The Mitchell Spectrograph is an integral-field
spectrograph composed of 246 fibers covering a
107\arcsec$\times$107\arcsec\ field of view with a one-third filling
factor. Each of the 246 fibers subtends $4\farcs2$ and they are assembled
in an array similar to Densepak \citep{bardenetal1998}. The Mitchell
Spectrograph has performed a very successful search for Ly$\alpha$
emitters \citep{adamsetal2011,finkelsteinetal2011,blancetal2011} and
has become a highly productive tool to study spatially resolved
kinematics and stellar populations in nearby galaxies
\citep{blancetal2009,yoachimetal2010,murphyetal2011,adamsetal2012}.

We use the low-resolution (R $\approx$ 850) blue setting of the Mitchell
Spectrograph. Our wavelength range spans 3550-5850~\AA\ with an average
spectral resolution of 5~\AA\ FWHM. This resolution delivers a
dispersion of $\sim$~1.1~\AA\ pixel$^{-1}$ and corresponds to
\sigmastar~$\approx 150$~\kms\ at 4300~\AA, our bluest Lick index.
Each galaxy was observed for a total of $\sim 2$ hours on source with
one-third of the time spent at each of three dither positions to fill
the field of view.  Initial data reduction is accomplished using the
custom code Vaccine \citep{adamsetal2011,murphyetal2011}. The details
of our data reduction are described in \citet{greeneetal2012} and
\citet{murphyetal2011}, so we repeat only a brief overview for
completeness here.

Initial overscan and bias subtraction are performed first on all
science and calibration frames. Twilight flats are used to construct a
trace for each fiber, which takes into account curvature in the
spatial direction, following \citet{kelson2003} to avoid interpolation
and thus correlated errors.  A wavelength solution is derived for each
fiber based on arcs taken both at the start and end of the night using
a fourth-order polynomial.  The typical residual variations about
this best-fit fourth-order polynomial are between 0.05 and 0.1~\AA\
depending on the night. The flat field is constructed from twilight
flats, with the solar spectrum modeled and removed.  The flat field is
stable to $<0.1$ pixel for typical thermal variations in
the instrument of less than 5 degrees Celsius.  When the temperature
variation exceeded 5 C over a night, the reductions are split, with
the nearest calibration frame in temperature being used.  The flat
field is then applied to all of the science frames, and corrects
variations in the individual pixel response, in the relative
fiber-to-fiber variation, and in the cross-dispersion profile shape
for every fiber.

The sky is modeled using off-galaxy sky frames observed with a
sky-object-object-sky pattern, with ten minute exposure times on sky
and twenty minute object exposures. The sky nods are processed in the
same manner as the science frames described above. In general, each
sky nod is weighted equally, although in unstable conditions (clouds,
for instance) we experiment with different weighting schemes to
achieve an optimal sky subtraction.  Since the galaxies are fainter
than the sky in their outskirts, sky subtraction is a limiting factor
for us.  We quantify our uncertainties due to sky subtraction in \S
\ref{sec:Error}. Finally, cosmic rays are identified and masked.

We use software developed for the VENGA project \citep{blancetal2009,blancetal2013}
for flux calibration and final processing.  We observe flux
calibration stars each night using a six-point dither pattern and
derive a relative flux calibration in the standard way.  Then we use
tools developed by M.~Song, et al. (in preparation) to derive an
absolute flux calibration relative to the SDSS imaging.  
Synthetic photometry is used to scale the spectra to match the SDSS
$g-$band image of each field.  Finally, all fibers are interpolated
onto the same wavelength scale and combined.  

Since most of our galaxies have SDSS spectra, we can test the
wavelength dependence of the flux calibration by comparing the shape
of the spectrum in the central fiber of the Mitchell Spectrograph with
the SDSS spectrum.  We find $\lesssim 5\%$ agreement in nearly all cases,
with no more than $\sim 10\%$ differences at worst.

\subsection{Effective radii}

In Paper I we adopted the SDSS model radius (the de Vaucouleurs fit)
as the effective radius ($R_e$).  While there is considerable evidence
that the shape of the light profile changes systematically with galaxy
mass \citep[e.g.,][]{caonetal1993,kormendyetal2009}, fitting the
galaxies with a fixed \sers\ index of four has the benefit that we are
less sensitive to both sky subtraction errors
\citep{mandelbaumetal2005,bernardietal2007} and to the detailed shape
of the light profile in the very faint wings
\citep[e.g.,][]{lacknergunn2012}.  In the effort to have a uniform
analysis, we have therefore adopted the effective radii published by
the SDSS. The galaxy NGC 6482 is not in the SDSS, and we have adopted
$R_e$ from NED in this case.  Below we will examine bins in physical
as well as $R_e$-scaled radii to mitigate uncertainties in the
measurement and meaning of $R_e$
\citep[e.g.,][]{kormendyetal2009,huangetal2013a}.

\subsection{Radial bins}

Individual spectra, with the exception of those at the very center of
the IFU, have inadequate signal for stellar population analysis.
Therefore, all of our analysis is performed on binned spectra.  We
utilize four binning schemes here, in all cases defining elliptical
annuli based on the axis ratio measured by the SDSS.  First, for the
maximum spatial resolution, we create bins with radial width of
4\arcsec, the width of an individual fiber.  Second, we make bins of
width $0.5 R_e$.  In Paper I, these two binning schemes were nearly
identical, given that most of the galaxies had effective radii of
$\sim 8$\arcsec.  However, in our larger sample, many of the galaxies
have effective radii of $10-25$\arcsec, requiring a finer binning
scheme.  In the end, our highest spatial resolution corresponds to
$0.2-0.5 R_e$ depending on the galaxy. Third, we make bins of width
$R_e$, from which we measure \sigmastar.  Fourth, we make bins with
fixed physical sizes of 0-15 kpc in 3 kpc increments.  For reference,
in Table \ref{tab:obs} we include the number of spectra that are
combined in the $1.5-2 R_e$ bin, and the surface brightness of each
galaxy at $2 R_e$.  We typically achieve a S/N of $>30$ per pixel at a
surface brightness brighter than $r < 23$ mag arcsec$^{-2}$.

\section{Analysis}
\label{sec:Analysis}

As discussed in detail in Paper I, we will use Lick indices as a tool
to trace the stellar populations, rather than full spectral synthesis
models, given both the imperfect flux calibration of our data and the
difficulty in modeling the effects of abundance ratio changes
\citep[e.g.,][]{wortheyetal1994} self-consistently
\citep[e.g.,][]{gallazzietal2005}.  In addition, we construct coadded
spectra to increase the contrast in percent-level variations
around absorption features of interest.

\subsection{Equivalent widths, emission line corrections, 
and stellar population modeling}

As in \citet{greeneetal2012}, we use {\it lick\_ew}
\citep{gravesschiavon2008} to measure the Lick indices.  First,
however, we must correct for low-level emission that can fill in the
absorption lines and artificially lower their equivalent widths
(EWs). Low-EW emission from warm ionized gas is very common in the
centers of elliptical galaxies \citep{sarzietal2010,yanblanton2012}.
In particular, the emission from \hbeta\ is weak in all cases, but
even a 0.1~\AA\ error in the \hbeta\ EW can lead to errors of $\sim 2$
Gyr in the modeling \citep[e.g.,][]{schiavon2007}.

\begin{figure*}
\hskip 15mm
\includegraphics[scale=.70,angle=90]{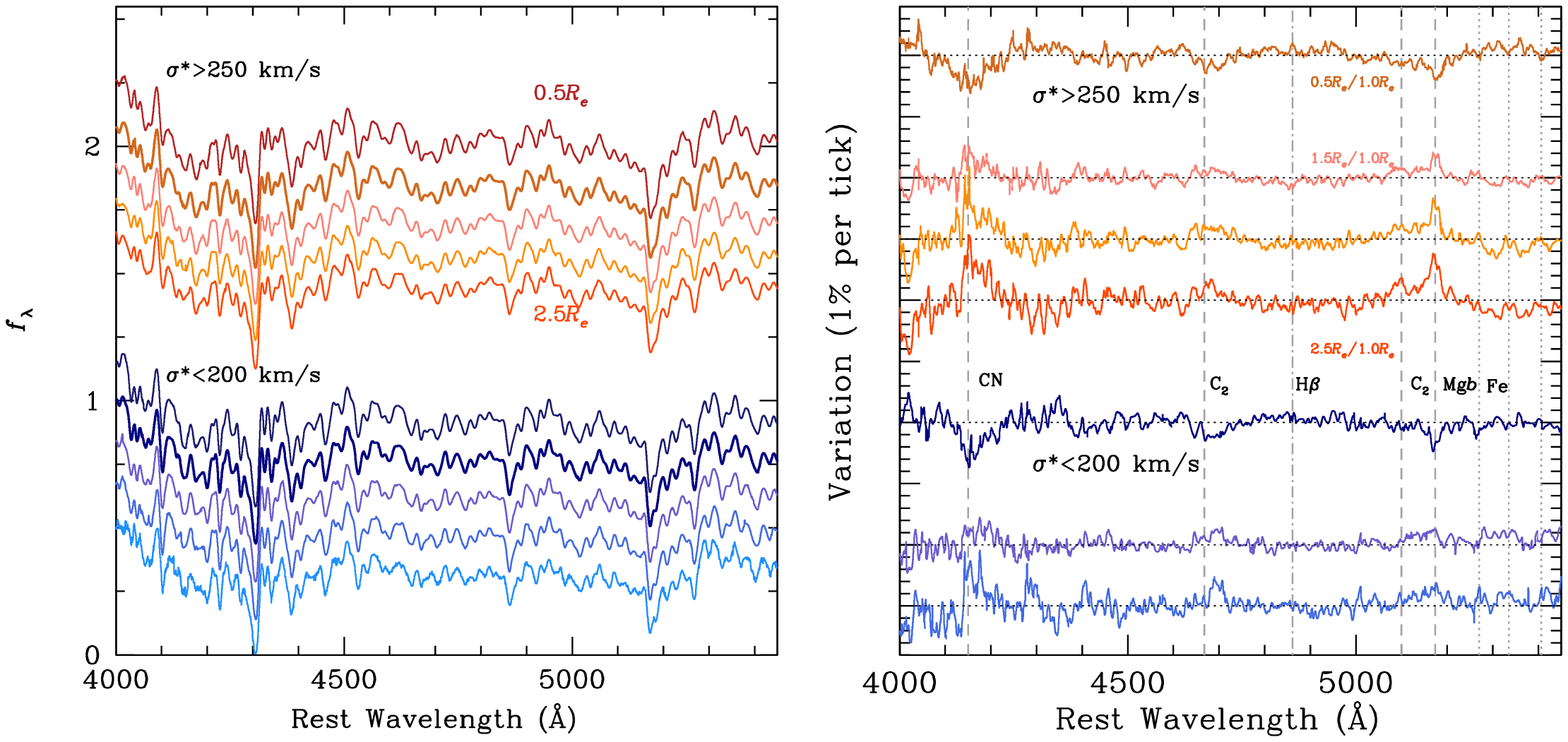}
\figcaption{{\it Left}: Composite spectra in radial bins of effective radius, from $0.5-2.5 R_e$. 
We have divided the sample into three \sigmastar\ bins, and show the largest 
($250 < $\sigmastar$< 300$~\kms; top five red spectra) and smallest dispersion 
($150 <$\sigmastar$\leq 200$~\kms; bottom five blue spectra) galaxies here.  All spectra 
have been normalized at 4500 \AA, and the offsets shown here are arbitrary.
{\it Right}: To highlight the differences between spectra in subsequent radial bins, 
we have divided each spectrum by the second ($0.5R_e < R < 1 R_e$) bin, 
shown in the same order 
as on the left.  Again, offsets are arbitrary but dotted lines denote unity, so we are 
seeing variations at the 1-5\% level in these spectra.  Vertical dashed lines highlight 
spectral bands that show large variations including Fe lines (dotted), 
molecular bands CN, C$_2$, and MgH (dashed) and H$\beta$ (dot-dashed).
The low-\sigmastar\ galaxies, being fainter,
have much lower surface brightness in their outer parts, and thus we only plot 
the residuals out to $2 R_e$, where we deem our results reliable.
\label{fig:coadd}
}
\end{figure*}

In Paper I, we utilized pPXF+GANDALF developed by M. Sarzi
\citep{sarzietal2006} and M. Cappellari \citep{cappellariemsellem2004}
to simultaneously model the stellar absorption and emission lines.
Here, instead, we fit each spectrum with an empirical template drawn
from the composite spectra of \citet{gravesetal2010}.  We then fit the
\oiii\ emission in the residual spectrum, and subtract both \oiii\ and
\hbeta, assuming that the \hbeta\ emission is 70\% of the \oiii\ flux
\citep[good to within a factor of
two,][]{trageretal2000a,gravesetal2007}.  In
addition, we fit Gaussians to residuals around strong sky lines at
5200 and 5460 \AA, and subtract them.

The strongest emission ($\gtrsim 0.2$ \AA\ in more than one radial
bin) is found in NGC 426, NGC 661, NGC 677, NGC 7509, NGC 7684, and
UGC 1382.  We found that our GANDALF results were sensitive to
template mismatch, and so adopted this iterative approach that gives
us more control over the templates but less control over the
line strengths. The results are reasonably consistent between the two
techniques for sources with \hbeta$>0.1$\AA\ in the GANDALF fits: 75\%
of these are detected in our iterative method, which returns \hbeta\
EWs that are $\sim 60\%$ weaker than those derived from GANDALF. 
In the future our goal is to refine the GANDALF measurements
using models with a range in [\alp/Fe]
\citep[e.g.,][]{coelhoetal2007,vazdekisetal2010,conroyetal2013}.

We then use {\it lick\_ew} \citep{gravesschiavon2008} on the
emission-line corrected spectra. This code corrects for both the
instrumental and intrinsic velocity dispersion, the latter measured
using pPXF. The indices are put onto a modified Lick system presented
by \citet{schiavon2007} based on flux-calibrated spectra.  In order to
demonstrate that we are on the same system, we compare the Lick
indices from the flux-calibrated SDSS spectra (the inner 3\arcsec)
with those from the central 4\arcsec\ fiber in our data, but we
exclude NGC 219, NGC 426, NGC 677, NGC 1267, and IC 301 from the
comparison due to the presence of bright stars near the nucleus. There
is no net offset between the two sets of indices in any case, with
$\langle ({\rm EW}_{\rm MS} - {\rm EW}_{\rm S})/{\rm EW}_{\rm S}
\rangle = 0.01 \pm 0.09$.  \hbeta\ and $\langle$Fe$\rangle$ each have
a scatter of $\sim 10\%$ while \mgb\ and the G-band each have a
scatter of $\sim 5\%$.

As in Paper I, we use the stellar population modeling code {\it
  EZ\_Ages} \citep{gravesschiavon2008} to convert the Lick indices
from the composite spectra to physical parameters (age, \feh,
[\alp/Fe]).  The code works on a hierarchy of index pairs, starting
with \hbeta\ and \feave, and iteratively solves for the age, abundance
and abundance ratios.  The models of \citet{schiavon2007} include
abundance ratio differences using the methodology of
\citet{trageretal2000a} and the response functions of
\citet{kornetal2005}. In our default runs, we utilize the
\alp-enhanced isochrone from \citet{salasnichetal2000} and the default
assumption that [O/Fe]$=0.5$.  We revisit this final assumption in \S
\ref{sec:Dstarform}.

\subsubsection{Uncertainties}
\label{sec:Error}

Our error budget is dominated by small errors in sky subtraction,
particularly when sky lines fall within the bands of the Lick index
measurements.  Thus, for each extracted spectrum, we generate eight
perturbed spectra with the fiducial sky subtraction scaled by $\pm 5,
4, 3, 2\%$ respectively.  We then run our entire procedure on these
perturbed spectra.  The final errors represent the spread in EWs
produced by these variations in sky subtraction.  As described in
Paper I, our sky subtraction uncertainty is unlikely to exceed 3\%, so
these error bars are conservative.

In Paper I, all of our targeted galaxies had similar sizes and
distances.  Therefore, at a given radial distance, we achieved similar
quality spectra for all galaxies.  In this sample we span a much wider
range in galaxy stellar mass, and reach our limiting surface
brightness at different radii for each galaxy.  At a certain point,
the Lick index measurements are no longer reliable.  To ensure some
consistency across all galaxies, we do not consider lick index
measurements from spectra with S/N per pixel $\leq 20$.

\subsection{Composite spectra}
\label{sec:Composite}

While measuring Lick indices is a very powerful technique at high S/N,
at the large radii that we are working systematic effects such as
small errors in sky subtraction and flux calibration can 
cause large uncertainties in the Lick indices measured from individual
objects. With our sample size, we benefit from
averaging over multiple galaxies at each radial bin.  The composite spectra will
suffer less from the vagaries of sky subtraction and flux calibration,
which occur at different wavelengths in each galaxy rest-frame
\citep[e.g.,][]{gravesetal2009,yan2011}.  We are able to examine
radial variations in the composite spectra at the percent level.

We first divide the galaxies into three bins of central \sigmastar\ of
$150 <$\sigmastar $< 200$, $200 <$ \sigmastar$< 250$, and $250 <$
\sigmastar$< 300$~\kms, since we know that the stellar population
properties are a strong function of \sigmastar\
\citep[e.g.,][]{wortheyetal1992,benderetal1993,trageretal2000b,gravesetal2009}.
Note that the bins would not change if we used \sigmastar\ within
$R_e$, but that in this way our bins are more consistent with the
large literature based on the SDSS measurements.  We combine spectra
that have already been emission-line--corrected. We interpolate the
rest-frame spectra onto the same wavelength grid and then smooth each
galaxy to the highest dispersion in the stack ($300$~\kms\ for the
high-dispersion bin and 200~\kms\ for the low-dispersion bin).
However, to increase the contrast, we will focus exclusively on the
highest and lowest dispersion bins here. We remove the continuum by
dividing each spectrum by a heavily smoothed version of itself. This
step simultaneously normalizes all spectra to the same level and
ensures that differences in continuum shape (whether real or due to
small errors in sky subtraction or flux calibration) do not impact the
final line strengths.  We then combine all pixels at each wavelength
using the biweight estimator, which should be robust even with limited
statistics \citep{beersetal1990}.  We experiment with multiplying the
composite spectrum by the median continuum before measuring indices,
but the changes to the Lick indices are negligible in all cases.

The composite spectra are shown in Figure \ref{fig:coadd}.  Again
we focus only the lowest and highest-dispersion bins, which contain 11
galaxies each (excluding NGC 219 due to severe night sky contamination 
and NGC 6482 due to contamination from a neighboring galaxy). 
The spectra are strikingly similar as a function of radius, so we use
ratio spectra to highlight the percent level radial variations (Figure
\ref{fig:coadd}, right).  Specifically, we divide each composite
spectrum, at each radial position, by the composite spectrum at 0.5-1
$R_e$.  We choose this radius, rather than the central bin, because as
we will see below the largest variance in spectral properties occurs
in the very center.  We will examine the radial trends in these
spectra in detail in \S \ref{sec:Radial}.  For now, we note only that
the strongest variations are seen in carbon, \mgb, and perhaps nitrogen.

To determine the level of variation in the composite spectra, we
generate 100 trial composite spectra by randomly drawing from the
total list of galaxies within that \sigmastar\ bin, with
replacement. We measure Lick indices from each of these 100 trial
spectra. We then assign errors on the Lick indices measured from the
stack derived to enclose 68\% of the Lick indices measured from the
100 trials.

\section{Radial Variations in Stellar Populations}
\label{sec:Radial}

We have extracted spectra out to 2.5$R_e$
for a sample of 33 local massive elliptical galaxies with stellar
velocity dispersions ranging from $150 < \sigmastar < 370$~\kms.  We
use the composite spectra to measure high fidelity 
radial trends in the stellar population properties, including
age, \feh, and detailed abundance ratio gradients.  From the gradients, 
we make inferences about when, where, and how the stars in
the outer parts were formed.  First, we confirm that we can recover
reliable Lick indices from our composite spectra (\S \ref{sec:lickrad}).
Second, we look at the dominant radial trends as revealed in ratios of
the composite spectra (\S \ref{sec:coaddrad}).  Third, and finally, we
present the radial trends in stellar populations (\S
\ref{sec:coadd_stellarpop}).

\subsection{Radial Trends in Lick Indices}
\label{sec:lickrad}

We first investigate the average radial profile in the Lick indices as
a function of radius in the high-dispersion ($250 <$\sigmastar$< 300$~\kms)
and low-dispersion ($150 <$\sigmastar$<200$~\kms) galaxies. We use Lick
indices measured from the composite spectra (\S \ref{sec:Composite})
and plot them as filled points as a function of radius in Figure
\ref{fig:radial}.  As a check on the composite spectra, we also
calculate the median indices at each radius from the individual galaxy
measurements, divided into the same high- and low-dispersion
groups. These median profiles are shown as lines in Figure
\ref{fig:radial}. The consistency between the composite and median
measurements gives us confidence in the measurements from our composite spectra.

We show radial profiles both as a function of radius
scaled to $R_e$ and in physical units of kpc.  Using bins scaled to
the effective radius is convenient when comparing galaxies of varying
size. However, there are two problems with using $R_e$-normalized
units. First, $R_e$ is difficult to measure, and becomes more so for
high-mass galaxies that have an extended low surface-brightness halo
\citep[e.g.,][]{kormendyetal2009}. Second, $R_e$ grows with cosmic
time.  Therefore, if we are searching for changes in stellar
populations that correspond to different epochs in galaxy growth, we
may want to look at physical as well as $R_e$-scaled radii.  

Interestingly, most of the variation in the Lick indices occurs within
the central $\sim 7$ kpc or $\sim 1.5 R_e$.  From photometric fitting,
\citet{huangetal2013a,huangetal2013b} find evidence for a distinct outer ($\sim 10$
kpc) component, perhaps formed via accretion. We see a tantalizing
hint that the index values are converging beyond $\gtrsim 2 R_e$
between the low- and high-dispersion galaxies. If indeed the central
compact regions of elliptical galaxies were formed in a very rapid
event at high redshift, with the outer parts accreted later
\citep[e.g.,][]{thomasetal2005,oseretal2010}, then we might be
reaching a radius where a large fraction of the mass has been accreted
from smaller galaxies.  As our next means of studying radial
stellar population trends, we examine the stacked spectra as a function of
radius.

\epsscale{.8}
\includegraphics[scale=.50,angle=0]{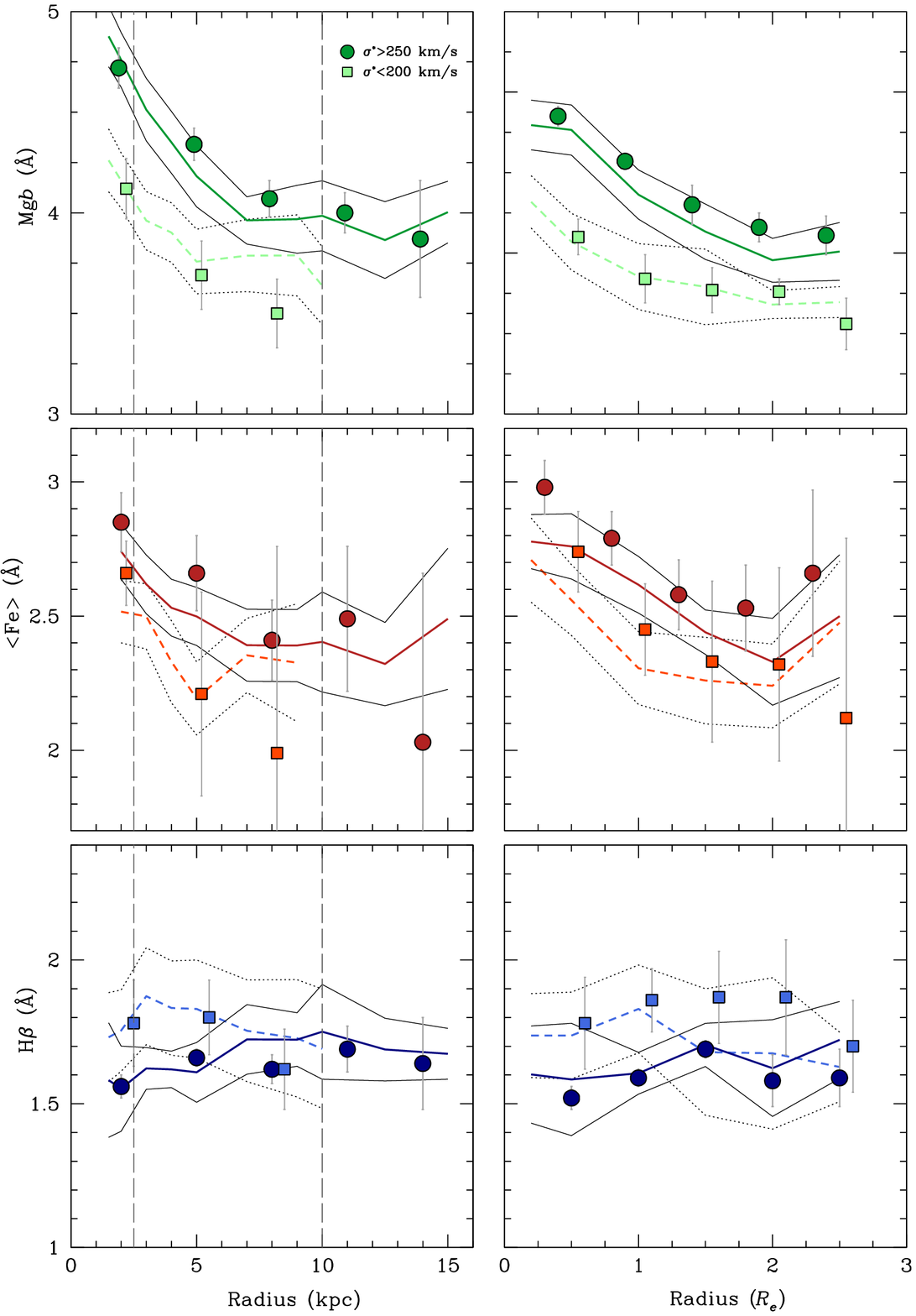}
\figcaption{We show the median and standard deviation in each 
index (\mgb, $\langle$Fe$\rangle$, and \hbeta) as a function of radius
(units of kpc on the left and $R_e$ on the right) 
for low-dispersion ($150 < $\sigmastar$< 200$~\kms; dotted lines) and 
high-dispersion galaxies ($250 < $\sigmastar$<300$~\kms; solid lines).  
Overplotted in points are the measurements from the composite spectra 
in the high-dispersion (filled circles) and low-dispersion (filled squares) galaxies. 
We also indicate the typical scales for the central and outer components 
of galaxies as measured in \citet[][grey vertical dashed lines]{huangetal2013a}.
\label{fig:radial}
}

\subsection{Radial Trends at the Percent Level}
\label{sec:coaddrad}

In addition to measuring more reliable Lick indices at large radius,
we examine percent-level variations directly from the stacked spectra.
While we do not derive any quantitative conclusions from this
exercise, it is nevertheless revealing to see where the largest
radial variance occurs in the spectra. We create
composite spectra for two groups of galaxies divided by high
(\sigmastar$=250-300$~\kms) and low (\sigmastar$=150-200$~\kms) stellar
velocity dispersion, and we add them radially in units of $R_e$
(Fig. \ref{fig:coadd}) and kpc (Fig. \ref{fig:coaddphys}).  Since the
largest variations occur within the central bin, we have made ratio
spectra (\S \ref{sec:Composite}) dividing by the composite spectrum in
the $0.5-1 R_e$ or the $3-6$ kpc bin respectively.  ``Emission''
features in the ratio spectra are manifestations of a declining EW
relative to $\sim R_e$, while ``absorption'' indicates increasing EW.

The most prominent radial changes are seen in the molecular bands CN
(the most prominent band is at 4150\AA), C$_2$ (the Swan band at
4668\AA, and another Swan band at $\sim 5100$\AA\ on the wing of the
\mgb\ line), and MgH (\mgb\ at 5200\AA).  While the
gentle decline in [Z/H] is known to be the dominant change in the
stellar populations as a function of radius (Paper I and references
therein), we see no strong variability in the pervasive atomic Fe
absorption features.  Also, both the CN and C$_2$ lines show strong
radial gradients, but we see no corresponding variation in the G-band
(CH) at 4300\AA.  According to \citet{tripiccobell1995}, the G-band is
more sensitive to microturbulent velocity (and thus effective
temperature) than abundances.  Therefore, we prefer to rely on the
C$_2$4668 index to infer carbon abundances
\citep[e.g.,][]{tripiccobell1995,wortheyetal1994,schiavon2007}. Finally,
we note that there also may be a shift in the centroid of the
C$_2$4668 index with radius, but higher S/N is needed to be certain.

The ratio spectra demonstrate very clearly what was already apparent
in the radial Lick index gradients (Figure \ref{fig:radial}).  The
largest spectral variations occur in the central regions, perhaps
reflecting a two-phase formation mode for these galaxies.  The strong
radial decrease in the C$_2$, CN, and MgH molecular bands reflects the
decreasing [Z/H]. At lower metallicity molecules do not form as
effectively.  However, as we show below, we also find compelling
evidence for true abundance ratio gradients as a function of radius,
particularly in [C/Fe].  

\subsection{Stellar population modeling}
\label{sec:coadd_stellarpop}

\begin{figure*}
\hskip 5mm
\includegraphics[scale=.98,angle=-90]{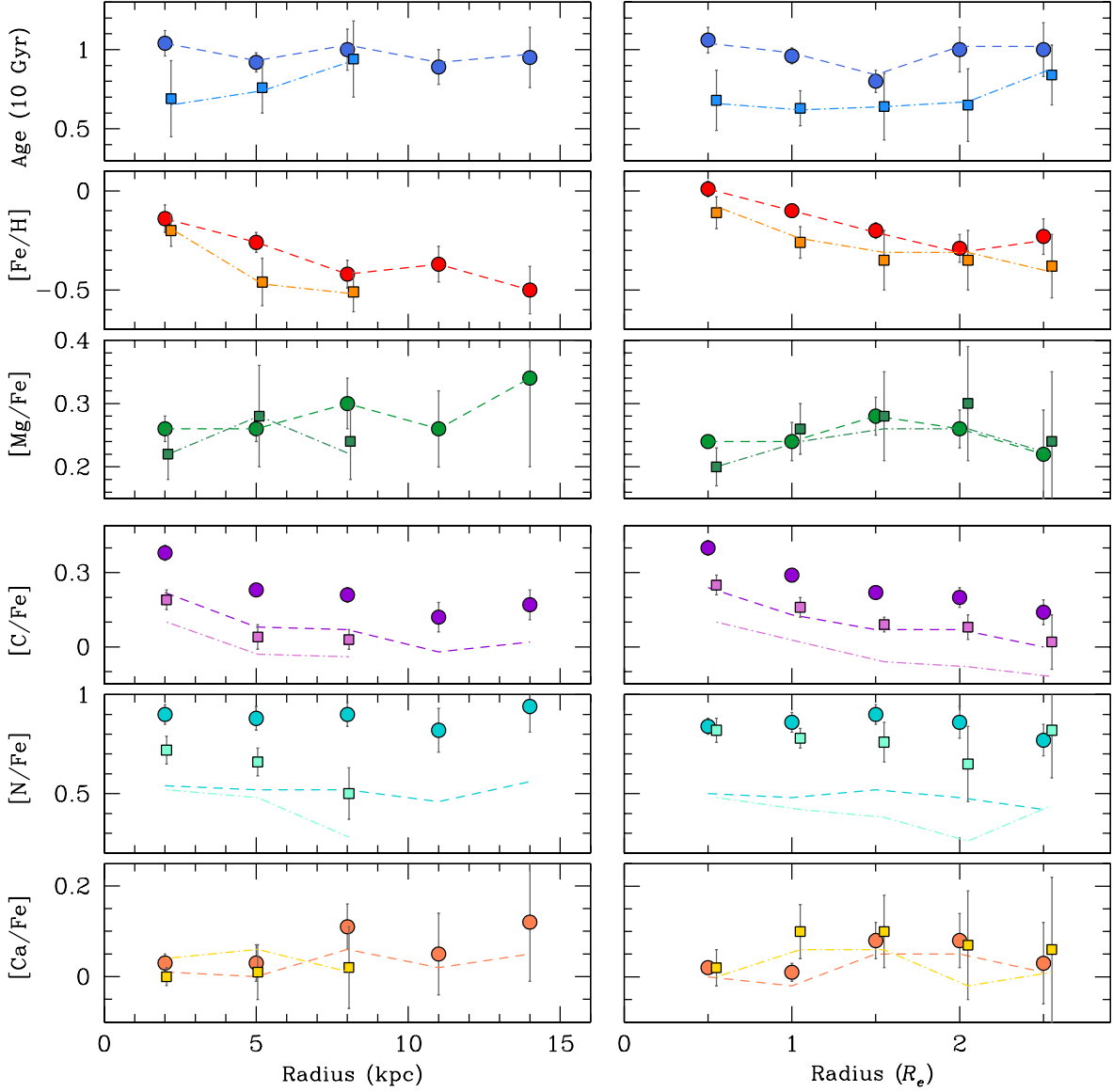}
\figcaption{Radial gradients in age, \feh, [Mg/Fe], [C/Fe], [N/Fe], and [Ca/Fe] as calculated by 
\emph{EZ\_Ages} from the Lick indices measured in the composite spectra. 
We show both the measurements for the high-dispersion (circles) and 
the low-dispersion (squares) galaxies as a function of $R$ in 
kpc (left) or $R/R_e$ (right).  At low dispersion, our observations do not reach beyond 
9 kpc. The lines show models of the same indices assuming 
[O/Fe]$=0.1$ rather than the default [O/Fe]$=0.5$, keeping [O/Fe] constant with 
radius in both cases (high-dispersion composite in dash and low-dispersion in dot-dash).
Note the decline with radius in \feh\ and [C/Fe] 
in contrast with the radially constant age, [Mg/Fe], [N/Fe], and [Ca/Fe].  
\label{fig:radialagefeafe}
}
\end{figure*}

Finally, we derive radial stellar population trends based on the
composite Lick index measurements presented in \S \ref{sec:lickrad}.
The resulting radial profiles in age, \feh, [Mg/Fe], [C/Fe], and
[N/Fe] are shown in Figure \ref{fig:radialagefeafe}.  The stellar
population trends inform us directly about when, where, and how rapidly 
the stars were formed and thus provide some 
clues to the assembly of the outer parts of massive elliptical galaxies.
Encouragingly, we recover well-known trends as a
function of velocity dispersion.  In their centers, higher-dispersion
galaxies are older, and have higher [Mg/Fe], [C/Fe] and [N/Fe] ratios
\citep[e.g.,][]{trageretal2000a,worthey2004,thomasetal2005,
  sanchezblazquezetal2006,gravesetal2007,
  smithetal2009,priceetal2011,johanssonetal2012,wortheyetal2013,conroyetal2013}.

Turning to the radial trends, we see first that the metallicity \feh\ drops
gently as a function of radius \citep[e.g.,][]{daviesetal1993}.  The
gradient $\Delta$\feh/$\Delta \,$log~$R_e$ has a similar slope for both the
high-dispersion and low-dispersion galaxies.  Previous work has found
interesting trends between \feh\ gradients and \sigmastar, but mostly
at lower \sigmastar\ than probed here
\citep[e.g.,][]{carolloetal1993,spolaoretal2010}.  The [Mg/Fe] ratio
stays high or even rises slightly for the low-dispersion galaxies
(Paper I). Only in age might we see tentative differences between high- and
low-dispersion galaxies: the low-dispersion galaxies show a weak
negative age gradient (get older) with radius, while the high
dispersion galaxies are old everywhere. New to
our analysis from Paper I, we also consider the [C/Fe], [N/Fe], and [Ca/Fe] ratios
as a function of radius.  We see a strikingly strong trend in the
radial decline of [C/Fe] with radius.  In contrast, [N/Fe] and [Ca/Fe]
more closely track the behavior of [Mg/Fe] and remain constant with
$R$.

As we already noted in regard to the Lick index gradients in \S
\ref{sec:lickrad}, the stellar population properties of the
high-dispersion and low-dispersion galaxies begin to converge beyond
$\sim 1.5 R_e$.  At large radius, the typical stars in all bins are
old, and have high \alp-abundances and low \feh\ \citep[a trend seen
in spiral bulges as well,][]{jablonkaetal2007}.  If galaxies are built
in two phases, then we might expect the largest variations in
properties to occur in their centers, where the formation timescales
and metal retention depend on the depth of the potential well of the
final galaxy.  

\vbox{
\vskip 20mm
\includegraphics[scale=.45,angle=0]{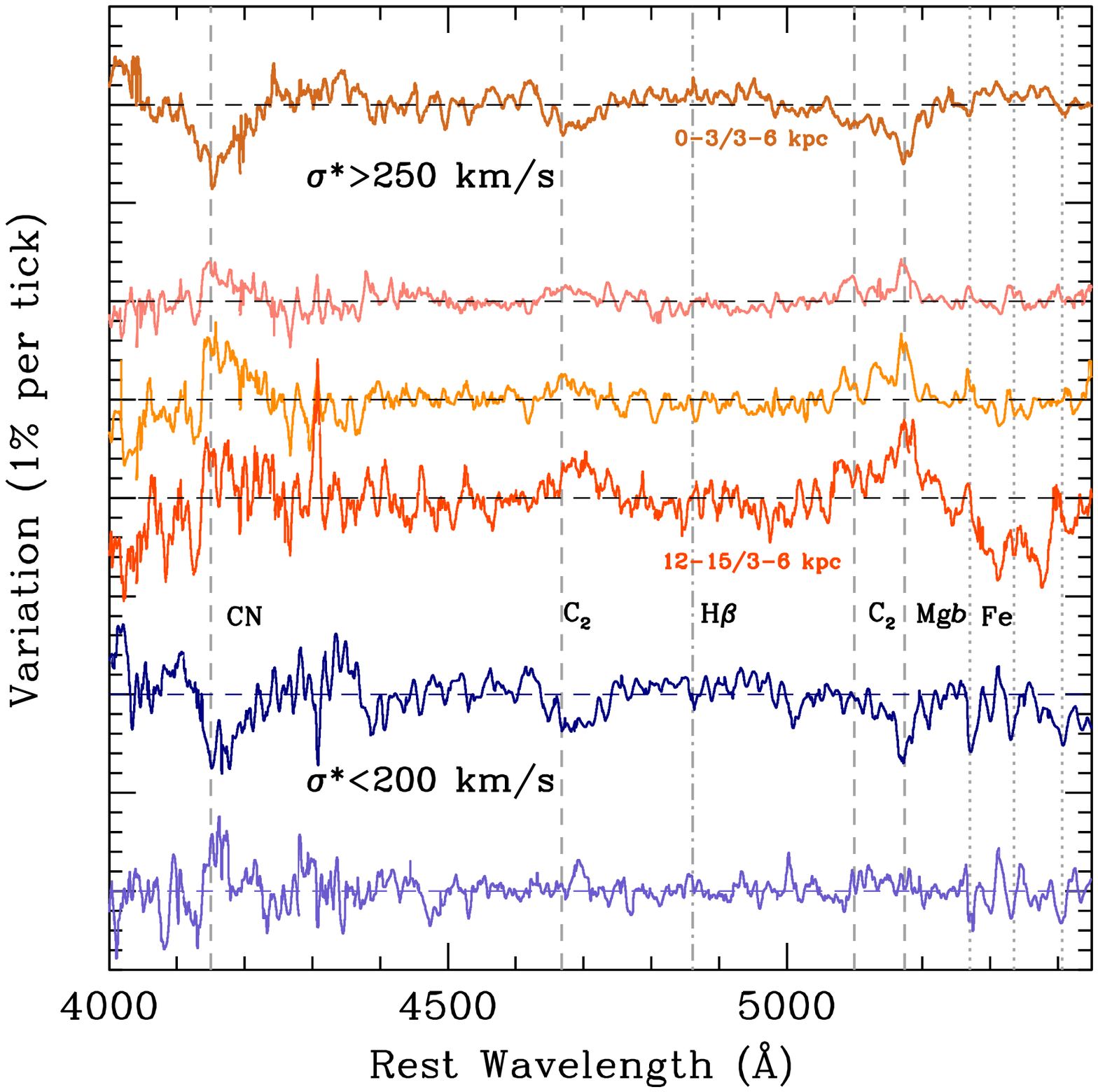}
}
\figcaption{Ratio spectra as in Figure \ref{fig:coadd} (right).  
Here, each radial bin has a fixed physical size of 3 kpc
(the 5 bins range from $0-15$ kpc) and we show the ratio with the 
$3-6$ kpc bin.   We cannot 
reach further than the $6-9$ kpc bin for the low-dispersion stack. 
Offsets are arbitrary but dotted lines denote unity, so we are 
seeing variations at the 1-5\% level in these spectra.  Vertical dashed lines highlight 
spectral bands that show large variations including Fe lines (dotted), 
molecular bands CN, C$_2$, and MgH (dashed) and H$\beta$ (dot-dashed).
\label{fig:coaddphys}
}
\vskip 5mm

\noindent
In contrast, if stars at large radius were accreted
from smaller systems that formed their stars rapidly and early
\citep[e.g.,][]{oseretal2010}, we expect stars in the outskirts to be
uniformly old, \alp-enhanced, and metal-poor, as we find for our
galaxies.

We note three caveats in interpreting the more detailed abundance
ratios, such as [C/Fe].  First, our CN measurements are potentially
problematic, due to the steep continuum shape around the 4000\AA\
break and the difficulties of accurate flux calibration and sky
subtraction at the blue edge of our spectra.  The [N/Fe] and [Ca/Fe]
abundances, because they depend on both CN and the C$_2$ measurements,
are both impacted by the uncertainty in measuring CN. However, the CN
index measured from our central spectra shows no systematic offset
from the CN index measured from the SDSS spectra, with (CN$_{\rm
  center}$ - CN$_{\rm SDSS}$)/CN$_{\rm SDSS} = 0.07 \pm 0.2$. 
This agreement gives us confidence that our CN index measurement is not 
driven by systematics in our reductions.

Second, we do not directly model [O/Fe], but the assumed oxygen
abundance directly impacts the [N/Fe] and [C/Fe] values
\citep{gravesetal2007}. Because most of the C is locked up in CO, a
slightly supersolar [C/O] leads to a large increase in the strength of
the C$_2$ Swan bands \citep[see
also][]{tripiccobell1995,kornetal2005}.  To bracket the uncertainty in
[O/Fe], we run a second set of stellar population models with
[O/Fe]$=0.1$ (rather than the default [O/Fe]$=0.5$; dotted lines in
Fig. \ref{fig:radialagefeafe}).  We make the reasonable assumption
that as an \alp\ element [O/Fe] tracks [Mg/Fe], as has been seen in
recent studies of elliptical galaxy centers.  \citet{conroyetal2013}
model oxygen abundance in SDSS galaxies using full spectral fitting
and find that the O/Mg ratio is constant to within 0.05 dex. Likewise,
\citet{johanssonetal2012} find O/Mg$\sim 1$ for all \sigmastar.  Thus,
adopting a range of [O/Mg] spanning $\pm 0.2$ dex should bracket the
range of allowed [C/Fe], [N/Fe], and [Ca/Fe].  Figure
\ref{fig:radialagefeafe} shows that as expected, only [C/Fe] and
[N/Fe] are strongly impacted by different values of [O/Fe].  At lower [O/Fe],
the absolute value of [C/Fe] and [N/Fe] are both correspondingly
lower, and we note that a value of [N/Fe]$\approx 0.5$ dex is
consistent with values reported for the most massive SDSS galaxies
\citep[][]{johanssonetal2012,conroyetal2013}. On the other hand, the
radial behavior of C and N relative to each other are not impacted by
the overall O abundance.

Third, we note that the models do not include carbon stars
\citep{schiavon2007}.  However, the incidence of carbon stars is low
at the metallicities considered here, and then increases strongly at
yet lower metallicity
\citep[e.g.,][]{blancomccarthy1983,groenewegen1999,mouhcinelancon2003}.
Thus, we do not believe carbon stars can be dominating the observed
[C/Fe] trends.

\subsection{Summary}

We find that stellar population gradients are strongest within $\sim
R_e$, while at larger radius the gradients begin to
flatten. Furthermore, the differences between the high- and
low-dispersion galaxies in terms of age and \alp-abundance decrease
with $R$; at all dispersions, stars at large radius are old,
\alp-enhanced, and relatively metal poor. Note that we include
  S0s in this analysis; we do not yet have a large enough sample to
  separate them, but will do so in future work.  Finally, we infer
strong negative gradients in [C/Fe] with radius, while the [N/Fe]
abundances are high (at least [N/Fe]$\approx 0.5$) and flat.  We now
discuss the ramifications of our results for the formation histories
of massive elliptical galaxies.

\section{Discussion}
\label{sec:Discussion}

If elliptical galaxies are built from the inside-out, with an early
dissipational phase making a compact ($\sim 2$ kpc) central component
followed by late-time accretion of satellites
\citep[e.g.,][]{oseretal2010}, we expect to see the imprint of
that process in the stellar populations as a function of radius.  The
detailed stellar population properties provide clues about when the stars 
formed, how quickly they were formed, and the depth of the potential 
well that they formed in. 

\subsection{Star Formation and Metal Production}
\label{sec:Dstarform}

Each stellar population property shown in Figure
\ref{fig:radialagefeafe} tells us something about the provenance of
these stars.  The stellar age obviously provides one important clue.
The metallicity presumably maps onto the depth of the potential well
in which the star formed \citep[e.g.,][]{larson1974}.  Finally,
[\alp/Fe] and the more detailed abundance ratios we consider here
reflect how rapidly the stellar population was formed.  High [\alp/Fe]
indicates a preponderance of Type II supernova relative to Type Ia,
and thus rapid formation timescales
\citep[e.g.,][]{matteuccigreggio1986}.  Finally, although their origins 
are less clear \citep[e.g.,][]{renzinivoli1981,cescuttietal2009}, [C/Fe] and [N/Fe]
provide complementary information about star formation timescales, as 
they are likely produced at least partially in intermediate-mass stars
\citep[e.g.,][]{gravesetal2007,johanssonetal2012}.  

To a large extent, the stellar populations in the centers of
elliptical galaxies scale with the stellar velocity dispersion.
Luminosity-weighted mean age, total metallicity [Z/H] and formation
timescale [\alp/Fe] all correlate with \sigmastar\
\citep[e.g.,][]{wortheyetal1994,trageretal2000a,gravesetal2007}.  The
more detailed abundance ratio patterns are likewise monotonic
functions of \sigmastar\ \citep[e.g.,][]{trageretal2000a,
  gravesetal2007,conroyetal2013,wortheyetal2013}.

As we discussed in Paper I, the stars at large radius do not
share the detailed stellar population properties of the stars at the
center of any present-day galaxy. The stellar ages are old and [Fe/H] is
0.3-0.4 dex subsolar. However, despite the low metallicity, 
the [\alp/Fe] abundance ratios are high \citep[e.g.,][Paper I]{spolaoretal2010}.  The
low \feh\ values are best explained by formation in shallow potential wells, either
of small galaxies as advocated in the two-phase model of galaxy
formation, or the outer parts of big galaxies.  However, given the old
ages and high [\alp/Fe] ratios, these stars must have formed early and
over short timescales. To estimate how rapidly, we adopt the scaling from
\citet{thomasetal2005} between [\alp/Fe] and star formation timescale,
which is based on a simple closed-box model. Assuming that [O/Fe]
tracks [Mg/Fe], we find a timescale of 250 Myr for [Mg/Fe]$\approx
0.3$. 

Likewise, the ratio of C/N apparently falls at large radius, while it
remains constant in the centers of elliptical galaxies.  As we argued
above, while the absolute values of carbon and nitrogen are uncertain
due to the unknown [O/Fe], their ratio as a function of radius
should be robust.  We will first discuss the radial behavior of
[C/Fe], and then turn to the more puzzling [N/Fe] trends.

Carbon is made in the triple alpha process in intermediate mass
\citep[$1-8$~\msun; e.g.,][]{renzinivoli1981} stars.  Massive,
metal-rich stars also release significant C through stellar winds
\citep[e.g.,][]{maeder1992}.  The rising [C/Fe] seen in the centers of
massive elliptical galaxies, where the star formation timescales are
short, seems to require that a significant fraction of the carbon
actually is made in massive stars
\citep[e.g.,][]{gravesetal2007}. Carbon from massive stars is also
needed to explain the correlation between the C/O and O/H ratios
observed in \hii\ regions and individual Milky Way stars
\citep[e.g.,][]{cescuttietal2009,garnettetal1999}.  Yield calculations
find that massive metal-rich stars can provide enough carbon to
explain the Galactic observations
\citep{maeder1992,marigoetal1998,henryetal2000}. Based on their high
[\alp/Fe], stars at large radius were likely formed rapidly, so the
declining [C/Fe] presumably reflects the declining carbon yields from
lower-metallicity massive stars, and supports our inference of rapid
star formation timescales.  

Nitrogen is more complicated. We must explain both the very
super-solar [N/Fe] values that we observe (at least [N/Fe]$\sim 0.5$)
and the falling C/N with radius. Nitrogen is produced in
intermediate-mass stars as part of the CNO cycle
\citep[e.g.,][]{matteucci1986}, but C is required before N can be
produced. The C is either synthesized in the star through the triple
alpha process (``primary'') or present in the star to begin with
(``secondary''). While some primary N is required from
low-metallicity massive stars to explain the floor in N abundances
seen in \ion{H}{2} regions
\citep[e.g.,][]{izotovthuan1999,meynetmaeder2002}, the yields from
massive stars alone are not high enough to produce the super-solar 
enrichment that we observe.  As
pointed out in \citet{johanssonetal2012}, high [N/Fe] ratios then
provide a lower limit on the star formation timescale of at least a
few Myr, the lifetimes of intermediate-mass stars.
Therefore, the star formation timescales inferred from [\alp/Fe] of 250 Myr are
consistent with the timescales required by [N/Fe].

Since synthesizing N requires existing C \citep{henryetal2000}, it is
possible that the high [N/Fe] results from the supersolar [C/Fe].
However, we would not expect [N/Fe] to remain
high at large radius where [C/Fe] is falling. Longer star formation
timescales could provide higher [N/Fe] ratios, but presumably would
flatten [C/Fe] as well. It could be that N is more effectively
released by lower metallicity stars, although we do not have a proposed
mechanism here. A changing IMF, in particular a bottom-heavy IMF as
have been invoked for the most massive elliptical galaxies
\citep{conroyvandokkum2012,thomasetal2011,duttonetal2012} cannot help,
since the C is made by massive stars. We do note that the nucleosynthis 
of N is a puzzle in globular clusters as well \citep[e.g.,][]{cohenetal2005}, 
and it may be that these problems have a similar solution. Intriguingly,
an ultra-compact dwarf in Virgo recently discussed by 
\citet{straderetal2013} also appears to have solar carbon abundances 
but supersolar [N/Fe]$\sim 0.6$ dex, 
perhaps providing an additional link between small stripped galaxies 
and elliptical stellar halos. Finally, there is always the
possibility that our [N/Fe] values are biased high due to
uncertainties in the CN measurements, although as mentioned above, we
find no systematic differences between our measurements and the SDSS
measurements. We have no clean explanation for the observed radial
decrease in C/N.

The conclusions for galaxy assembly are as follows.  We have seen that
stellar populations at radii $\gtrsim 2 R_e$ do not look like those
found at the centers of {\it any} elliptical galaxies today (Paper I,
and this work).  According to our stellar population modeling, stars
at $\sim 2 R_e$ are similar at all \sigmastar\ (see how they converge
at large radius in Fig. \ref{fig:afe_feh}). The typical star is old
($\sim 10$ Gyr), relatively metal poor (\feh$\approx -0.5$), and
\alp-enhanced ([Mg/Fe]$\approx 0.3$). We infer that stars at large
radius are formed at $z \approx 1.5-2$ in shallow potential wells over
$\sim 250$ Myr timescales. Declining [C/Fe] ratios support this
picture: with rapid star formation timescales, the declining C yields
from massive stars at low-metallicity leads to a decline in this ratio
with radius.  On the other hand, the high and flat [N/Fe] ratios are
unexpected, but possibly also seen in stripped dwarf galaxies.

\subsection{Galaxy Assembly}
\label{sec:Dassemble}

Given our observed radial trends in stellar population properties, we
return to the assembly history of massive elliptical galaxies.  Based
on the striking average size growth of elliptical galaxies between
redshift two and the present, numerous papers have proposed a
two-phase model for their growth
\citep[e.g.,][]{naabetal2009,bezansonetal2009,oseretal2010,oseretal2012,hilzetal2012}.
Our observations are consistent with this picture. The stars are
converging towards similar properties at large radius independent of
\sigmastar.  The stellar populations at large radius demand that the
accreted galaxies must be small (to explain the low \feh) and form
early and rapidly (to explain the old ages and high [\alp/Fe]).
Simulations make similar predictions \citep{oseretal2010}.

While the stars in the outskirts of massive ellipticals do not
resemble the centers of any elliptical galaxies today, there must be
present-day stars with similar metallicities and abundance patterns
that did not end up in massive elliptical galaxy outskirts.  In Figure
\ref{fig:afe_feh}, we show that the average star in the outskirts of
our ellipticals resembles thick-disk Milky Way stars.  Perhaps the
notion of a thick disk is outdated. Instead, we can say that stars in
elliptical galaxy outskirts are similar to the luminosity weighted
mean star in the Milky Way disk \citep[][Bovy private
communication]{bovyetal2012a,bovyetal2012b}.  Since stars form in
disks, and low-density disks are easier to disrupt than centrally
concentrated ellipticals, we suggest that the outer parts of our
elliptical galaxies were built by the shredding of disky galaxies at
early times \citep[e.g.,][]{toftetal2007,conseliceetal2011}.  

\vbox{
\vskip 15mm
\includegraphics[scale=.45,angle=0]{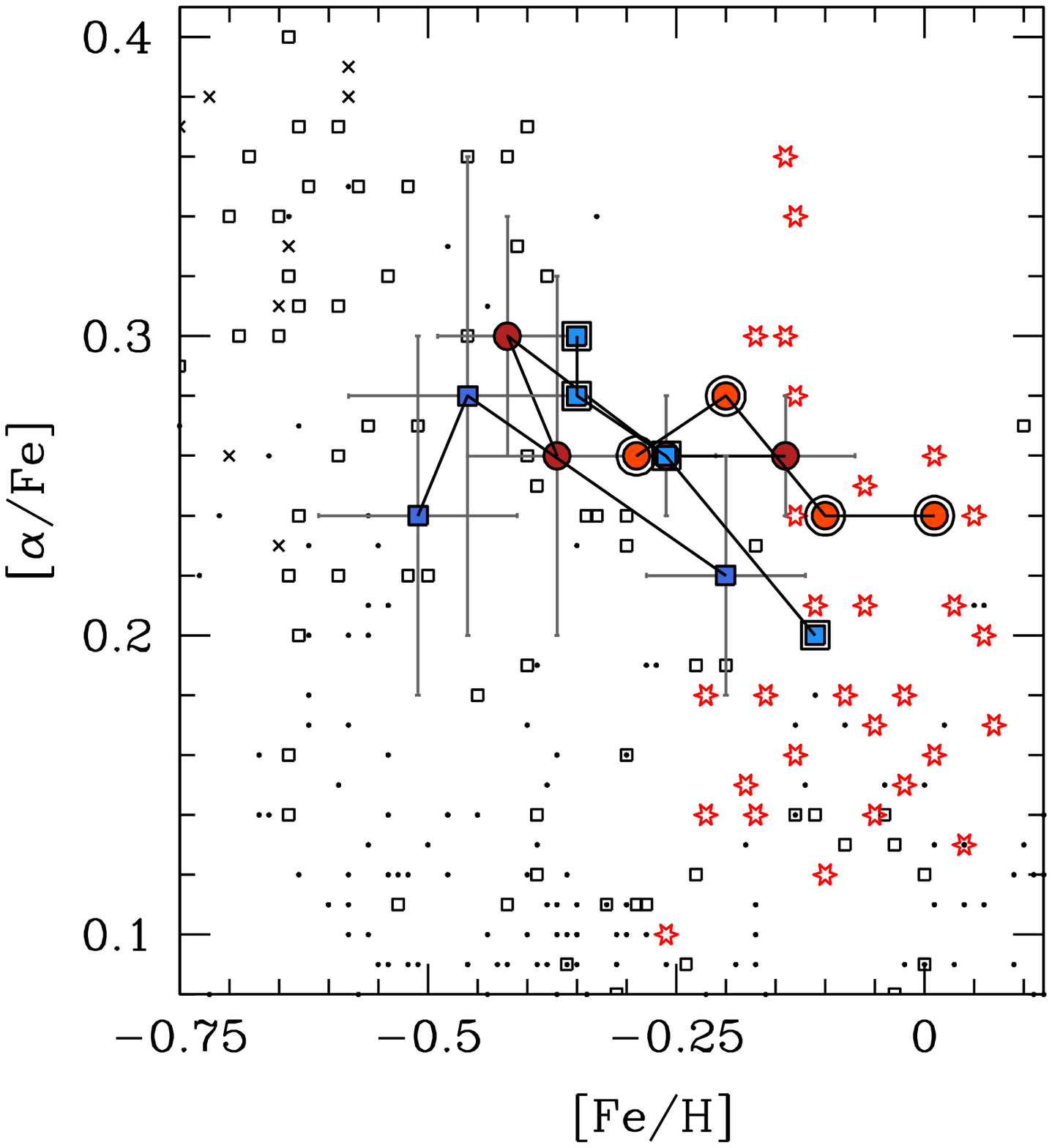}
}
\figcaption{
Movement of the stellar populations in the composite spectra through [\alp/Fe]-\feh\
space based on {\it EZ\_Ages} modeling of the Lick indices.  
The high-dispersion ($250 < $\sigmastar$<300$~\kms; filled red
circles) and low-dispersion galaxies ($150 < $\sigmastar$<200$~\kms; filled blue
squares) are shown both in kpc bins (filled symbols) and $R_e$ bins (double circles or 
squares). Both dispersion bins move systematically towards lower \feh\ and higher [\alp/Fe]
at larger radius, as indicated by the connecting lines.  For
reference, we also show the centers of SDSS galaxies from
\citet{gravesetal2010} in red open stars.  We compare the abundance
ratios and metallicities in our stellar halos with Milky Way stars
from \citet{vennetal2004}, including thin disk ({\it black dots}),
thick disk ({\it small black open squares}), and halo ({\it small
  black stars}) stars. 
\label{fig:afe_feh}
}
\vskip 5mm

\noindent
Based on the stellar populations, similar stars, formed at $z \approx
1.5-2$ in disks but never accreted by a more massive halo, form the
thick disk components of present-day spiral galaxies (see
Fig. \ref{fig:afe_feh}).  Our observations are thus consistent with a
two-phase model for elliptical galaxy growth.

On the other hand, our observations alone do not require such a
model. Instead, the average size growth in the elliptical galaxy
population may occur as larger galaxies join the red sequence at later
times \citep{valentinuzzietal2010,cassataetal2011,newmanetal2012,
  carolloetal2013,barroetal2013}, consistent with the observation that
at a given velocity dispersion, more diffuse galaxies quenched later
\citep{gravesetal2010}.  If stars are formed in situ at large radius,
followed by global star formation quenching, then we might expect
stellar population gradients as predicted by classic monolithic
collapse models \citep[e.g.,][]{larson1974,carlberg1984}.  
Monolithic collapse naively predicts very steep metallicity gradients
as a function of radius
\citep[e.g.,][]{larson1974,kobayashi2004}. However, the shallower
observed \feh\ gradients, as well as flat [\alp/Fe] gradients, can be
reproduced in monolithic scenarios by varying the star formation
efficiency and gas inflow rate \citep{pipinoetal2010}.  In situ
formation at large radius also more naturally explains the observation
that \mgb\ EW correlates strongly with local escape velocity
\citep[e.g.,][]{franxillingworth1990,weijmansetal2009,scottetal2009,scottetal2013}.

Both assembly paths (two-phase and all in situ) likely occur in nature
\citep[e.g.,][]{faberetal2007,toftetal2007,bundyetal2010,
  szomoruetal2012,barroetal2013}.  In the future, we hope that
combining our stellar population measurements with kinematics may
provide further clues to the level of dissipation involved in forming
stars at large radius \citep[e.g.,][]{hilzetal2012}.  We will grow
more sensitive to any such trends as our survey continues
(Raskutti et al. in preparation).

\subsection{Compact High-Dispersion Galaxies}

Eventually, we would like to examine the radial behavior of elliptical
galaxies as a function of their fundamental plane location (or
compactness) as in \citet{gravesetal2009,gravesetal2010}.  In the
meantime, we can look at one spectacular outlier in our sample: NGC1270.
This galaxy has a half-light radius of $\sim 2$~kpc, and a central
stellar velocity dispersion of $\sim 370$~\kms.
\citet{vandenboschetal2012} highlight NGC 1277 as the prototype of
these compact high-dispersion galaxies, which also appear to contain
very massive ($> 10^{10}$~\msun) supermassive black holes.  van den
Bosch and collaborators have found a handful of galaxies like NGC
1277, including NGC 1270 \citep[for low-mass versions
see][]{trujilloetal2009,jiangetal2012, ferremateuetal2012}. They have
disky light profiles and are rotationally dominated at large radius.

NGC 1270 appears to be an extreme example of a compact galaxy that
formed at high redshift, and then grew no more. With a central age of
$\sim 11$ Gyr, it is one of the oldest galaxies in the sample (Paper
I), has a very high central \mgb\ EW, and shows no gradients as a
function of radius ($\Delta$ log Age/$\Delta$ log $R_e =−0.20 \pm 0.13$). As 
far as we can tell, NGC 1270 follows the \mgb-\sigmastar\ relation 
(Paper I) but the statistics are very limited for velocity dispersions so high.
Based on color gradients and spectroscopy respectively, NGC 1277 and
SDSS J151741.75$-$004217.6 also seem to fit this picture
\citep{vandenboschetal2012,laskeretal2013}. We will look in more
detail at the stellar population gradients for a larger set of the
entire van den Bosch sample (Yildrim et al.\ in preparation).

\section{Summary}
\label{sec:Summary}

We have presented stellar population information as a function of
radius for 33 massive elliptical galaxies with stellar dispersions
\sigmastar$=150-370$~\kms. In addition to the well-known gentle
decline in \feh\ with radius, we find that the \alp-abundance ratios,
as traced by [Mg/Fe], are constant with radius (even perhaps rising
slightly at \sigmastar$< 200$~\kms), implying rapid star formation
timescales at all radii.  The declining [C/Fe] radial trends, we believe, 
reflect declining carbon yields from metal-poor massive stars, in line with 
the rapid star formation timescales.  However, 
we do not have a complete understanding of the rising N/C ratio that 
we observe in the outer parts.

At large radius, the stellar populations of the stars depend very
weakly on central \sigmastar; in general they have [Fe/H]$\approx
-0.4$, [Mg/Fe]$\approx 0.3$, and age$\approx 10$ Gyr.  These are
properties akin to stars in our own Milky Way thick disk, lending
credence to the idea that the outer parts of massive elliptical
galaxies comprise shredded disky galaxies whose star formation was
truncated at $z \approx 1.5-2$. The one very compact galaxy in our
sample has old ages and no abundance ratio gradients, apparently
because it never accreted any gas or stars at late times.

We should note that we only reach $\sim 2.5 R_e$ in this study. There
is evidence from a small number of integrated light studies
\citep[e.g.,][]{coccatoetal2010} and resolved studies
\citep[e.g.,][]{harrisetal2007} that the stellar populations may
change more dramatically at yet larger radius.  In the case of 
Coccato et al., they are also studying a brightest cluster galaxy 
NGC 4889, so it is difficult to know whether the declining [\alp/Fe] 
they observe is due to larger radius or the special location of the central 
galaxy deep in a cluster potential well. It is still extremely
observationally challenging to reach such large radius for more than a
handful of targets.

In the short term, we are in the process of doubling the current
sample.  With a larger sample, we hope to use the combination of
stellar populations and kinematics to pin down the formation paths of
individual elliptical galaxies. To derive meaningful stellar
  population measurements out to $2.5 R_e$ for individual galaxies we
  will utilize full spectral synthesis models with \alp\ dependence,
  to mitigate sensitivity to systematics such as sky subtraction
  \citep[e.g.,][]{coelhoetal2007,vazdekisetal2010,conroyetal2013}.
We will also look for differences in stellar population trends as a
function of environment and galaxy shape.  Finally, we look forward to
comparing our observations to more sophisticated cosmological models
that track abundance changes with time
\citep[e.g.,][]{tragersomerville2009,arrigonietal2010,pipinoetal2010,yatesetal2013}.

\acknowledgements
We thank A. Burrows, L. Coccato, J. Cohen, C. Conroy, L. Ho, G. Knapp, and
S. Trager for useful discussions about carbon. We thank A. Silverman
for suggesting we plot ratio spectra. J. D. Murphy and J. M. Comerford
are supported by Astronomy and Astrophysics Postdoctoral Fellowships
under awards NSF AST-1203057 and AST-1102525 respectively. J. E. Gunn
is partially supported by award NSF AST-0908368. This research has
made use of the NASA/IPAC Extragalactic Database (NED) which is
operated by the Jet Propulsion Laboratory, California Institute of
Technology, under contract with the National Aeronautics and Space
Administration.


\begin{thebibliography}{156}
\expandafter\ifx\csname natexlab\endcsname\relax\def\natexlab#1{#1}\fi

\bibitem[{{Adams} {et~al.}(2012){Adams}, {Gebhardt}, {Blanc}, {Fabricius},
  {Hill}, {Murphy}, {van den Bosch}, \& {van de Ven}}]{adamsetal2012}
{Adams}, J.~J., {Gebhardt}, K., {Blanc}, G.~A., {Fabricius}, M.~H., {Hill},
  G.~J., {Murphy}, J.~D., {van den Bosch}, R.~C.~E., \& {van de Ven}, G. 2012,
  \apj, 745, 92

\bibitem[{{Adams} {et~al.}(2011)}]{adamsetal2011}
{Adams}, J.~J., {et~al.} 2011, \apjs, 192, 5

\bibitem[{{Annibali} {et~al.}(2007){Annibali}, {Bressan}, {Rampazzo},
  {Zeilinger}, \& {Danese}}]{annibalietal2007}
{Annibali}, F., {Bressan}, A., {Rampazzo}, R., {Zeilinger}, W.~W., \& {Danese},
  L. 2007, \aap, 463, 455

\bibitem[{{Arrigoni} {et~al.}(2010){Arrigoni}, {Trager}, {Somerville}, \&
  {Gibson}}]{arrigonietal2010}
{Arrigoni}, M., {Trager}, S.~C., {Somerville}, R.~S., \& {Gibson}, B.~K. 2010,
  \mnras, 402, 173

\bibitem[{{Baes} {et~al.}(2007){Baes}, {Sil'chenko}, {Moiseev}, \&
  {Manakova}}]{baesetal2007}
{Baes}, M., {Sil'chenko}, O.~K., {Moiseev}, A.~V., \& {Manakova}, E.~A. 2007,
  \aap, 467, 991

\bibitem[{{Barden} {et~al.}(1998){Barden}, {Sawyer}, \&
  {Honeycutt}}]{bardenetal1998}
{Barden}, S.~C., {Sawyer}, D.~G., \& {Honeycutt}, R.~K. 1998, in Society of
  Photo-Optical Instrumentation Engineers (SPIE) Conference Series, Vol. 3355,
  Society of Photo-Optical Instrumentation Engineers (SPIE) Conference Series,
  ed. {S.~D'Odorico}, 892--899

\bibitem[{{Barro} {et~al.}(2013)}]{barroetal2013}
{Barro}, G., {et~al.} 2013, \apj, 765, 104

\bibitem[{{Beers} {et~al.}(1990){Beers}, {Flynn}, \&
  {Gebhardt}}]{beersetal1990}
{Beers}, T.~C., {Flynn}, K., \& {Gebhardt}, K. 1990, \aj, 100, 32

\bibitem[{{Bender} {et~al.}(1993){Bender}, {Burstein}, \&
  {Faber}}]{benderetal1993}
{Bender}, R., {Burstein}, D., \& {Faber}, S.~M. 1993, \apj, 411, 153

\bibitem[{{Bernardi} {et~al.}(2007){Bernardi}, {Hyde}, {Sheth}, {Miller}, \&
  {Nichol}}]{bernardietal2007}
{Bernardi}, M., {Hyde}, J.~B., {Sheth}, R.~K., {Miller}, C.~J., \& {Nichol},
  R.~C. 2007, \aj, 133, 1741

\bibitem[{{Bezanson} {et~al.}(2009){Bezanson}, {van Dokkum}, {Tal},
  {Marchesini}, {Kriek}, {Franx}, \& {Coppi}}]{bezansonetal2009}
{Bezanson}, R., {van Dokkum}, P.~G., {Tal}, T., {Marchesini}, D., {Kriek}, M.,
  {Franx}, M., \& {Coppi}, P. 2009, \apj, 697, 1290

\bibitem[{{Blanc} {et~al.}(2009){Blanc}, {Heiderman}, {Gebhardt}, {Evans}, \&
  {Adams}}]{blancetal2009}
{Blanc}, G.~A., {Heiderman}, A., {Gebhardt}, K., {Evans}, II, N.~J., \&
  {Adams}, J. 2009, \apj, 704, 842

\bibitem[{{Blanc} {et~al.}(2011)}]{blancetal2011}
{Blanc}, G.~A., {et~al.} 2011, \apj, 736, 31

\bibitem[{{Blanc} {et~al.}(2013)}]{blancetal2013}
---. 2013, \aj, accepted (ArXiv:1303.1552)

\bibitem[{{Blanco} \& {McCarthy}(1983)}]{blancomccarthy1983}
{Blanco}, V.~M., \& {McCarthy}, M.~F. 1983, \aj, 88, 1442

\bibitem[{{Bovy} {et~al.}(2012{\natexlab{a}}){Bovy}, {Rix}, \&
  {Hogg}}]{bovyetal2012a}
{Bovy}, J., {Rix}, H.-W., \& {Hogg}, D.~W. 2012{\natexlab{a}}, \apj, 751, 131

\bibitem[{{Bovy} {et~al.}(2012{\natexlab{b}}){Bovy}, {Rix}, {Liu}, {Hogg},
  {Beers}, \& {Lee}}]{bovyetal2012b}
{Bovy}, J., {Rix}, H.-W., {Liu}, C., {Hogg}, D.~W., {Beers}, T.~C., \& {Lee},
  Y.~S. 2012{\natexlab{b}}, \apj, 753, 148

\bibitem[{{Brough} {et~al.}(2007){Brough}, {Proctor}, {Forbes}, {Couch},
  {Collins}, {Burke}, \& {Mann}}]{broughetal2007}
{Brough}, S., {Proctor}, R., {Forbes}, D.~A., {Couch}, W.~J., {Collins}, C.~A.,
  {Burke}, D.~J., \& {Mann}, R.~G. 2007, \mnras, 378, 1507

\bibitem[{{Bundy} {et~al.}(2010)}]{bundyetal2010}
{Bundy}, K., {et~al.} 2010, \apj, 719, 1969

\bibitem[{{Caon} {et~al.}(1993){Caon}, {Capaccioli}, \&
  {D'Onofrio}}]{caonetal1993}
{Caon}, N., {Capaccioli}, M., \& {D'Onofrio}, M. 1993, \mnras, 265, 1013

\bibitem[{{Cappellari} \& {Emsellem}(2004)}]{cappellariemsellem2004}
{Cappellari}, M., \& {Emsellem}, E. 2004, \pasp, 116, 138

\bibitem[{{Cappellari} {et~al.}(2006)}]{cappellarietal2006}
{Cappellari}, M., {et~al.} 2006, \mnras, 366, 1126

\bibitem[{{Cappellari} {et~al.}(2009)}]{cappellarietal2009}
---. 2009, \apjl, 704, L34

\bibitem[{{Cappellari} {et~al.}(2012)}]{cappellarietal2012}
---. 2012, \nat, 484, 485

\bibitem[{{Carlberg}(1984)}]{carlberg1984}
{Carlberg}, R.~G. 1984, \apj, 286, 416

\bibitem[{{Carollo} \& {Danziger}(1994)}]{carollodanziger1994}
{Carollo}, C.~M., \& {Danziger}, I.~J. 1994, \mnras, 270, 743

\bibitem[{{Carollo} {et~al.}(1993){Carollo}, {Danziger}, \&
  {Buson}}]{carolloetal1993}
{Carollo}, C.~M., {Danziger}, I.~J., \& {Buson}, L. 1993, \mnras, 265, 553

\bibitem[{{Carollo} {et~al.}(2013)}]{carolloetal2013}
{Carollo}, C.~M., {et~al.} 2013, ArXiv e-prints

\bibitem[{{Cassata} {et~al.}(2010)}]{cassataetal2010}
{Cassata}, P., {et~al.} 2010, \apjl, 714, L79

\bibitem[{{Cassata} {et~al.}(2011)}]{cassataetal2011}
---. 2011, \apj, 743, 96

\bibitem[{{Cescutti} {et~al.}(2009){Cescutti}, {Matteucci}, {McWilliam}, \&
  {Chiappini}}]{cescuttietal2009}
{Cescutti}, G., {Matteucci}, F., {McWilliam}, A., \& {Chiappini}, C. 2009,
  \aap, 505, 605

\bibitem[{{Cimatti} {et~al.}(2008)}]{cimattietal2008}
{Cimatti}, A., {et~al.} 2008, \aap, 482, 21

\bibitem[{{Coccato} {et~al.}(2010){Coccato}, {Gerhard}, \&
  {Arnaboldi}}]{coccatoetal2010}
{Coccato}, L., {Gerhard}, O., \& {Arnaboldi}, M. 2010, \mnras, 407, L26

\bibitem[{{Coccato} {et~al.}(2011){Coccato}, {Gerhard}, {Arnaboldi}, \&
  {Ventimiglia}}]{coccatoetal2011}
{Coccato}, L., {Gerhard}, O., {Arnaboldi}, M., \& {Ventimiglia}, G. 2011, \aap,
  533, A138

\bibitem[{{Coelho} {et~al.}(2007){Coelho}, {Bruzual}, {Charlot}, {Weiss},
  {Barbuy}, \& {Ferguson}}]{coelhoetal2007}
{Coelho}, P., {Bruzual}, G., {Charlot}, S., {Weiss}, A., {Barbuy}, B., \&
  {Ferguson}, J.~W. 2007, \mnras, 382, 498

\bibitem[{{Cohen} {et~al.}(2005){Cohen}, {Briley}, \&
  {Stetson}}]{cohenetal2005}
{Cohen}, J.~G., {Briley}, M.~M., \& {Stetson}, P.~B. 2005, \aj, 130, 1177

\bibitem[{{Conroy} {et~al.}(2013){Conroy}, {Graves}, \& {van
  Dokkum}}]{conroyetal2013}
{Conroy}, C., {Graves}, G., \& {van Dokkum}, P. 2013, \apj, submitted
  (arXiv:1306.6629)

\bibitem[{{Conroy} \& {van Dokkum}(2012)}]{conroyvandokkum2012}
{Conroy}, C., \& {van Dokkum}, P.~G. 2012, \apj, 760, 71

\bibitem[{{Conselice} {et~al.}(2011)}]{conseliceetal2011}
{Conselice}, C.~J., {et~al.} 2011, \mnras, 417, 2770

\bibitem[{{Crnojevi{\'c}} {et~al.}(2013){Crnojevi{\'c}}, {Ferguson}, {Irwin},
  {Bernard}, {Arimoto}, {Jablonka}, \& {Kobayashi}}]{crnojevicetal2013}
{Crnojevi{\'c}}, D., {Ferguson}, A.~M.~N., {Irwin}, M.~J., {Bernard}, E.~J.,
  {Arimoto}, N., {Jablonka}, P., \& {Kobayashi}, C. 2013, \mnras, accepted
  (arXiv:1303.4736)

\bibitem[{{Daddi} {et~al.}(2005)}]{daddietal2005}
{Daddi}, E., {et~al.} 2005, \apj, 626, 680

\bibitem[{{Damjanov} {et~al.}(2009)}]{damjanovetal2009}
{Damjanov}, I., {et~al.} 2009, \apj, 695, 101

\bibitem[{{Davies} {et~al.}(1993){Davies}, {Sadler}, \&
  {Peletier}}]{daviesetal1993}
{Davies}, R.~L., {Sadler}, E.~M., \& {Peletier}, R.~F. 1993, \mnras, 262, 650

\bibitem[{{de Vaucouleurs} {et~al.}(1991){de Vaucouleurs}, {de Vaucouleurs},
  {Corwin}, {Buta}, {Paturel}, \& {Fouqu{\'e}}}]{devaucouleursetal1991}
{de Vaucouleurs}, G., {de Vaucouleurs}, A., {Corwin}, Jr., H.~G., {Buta},
  R.~J., {Paturel}, G., \& {Fouqu{\'e}}, P. 1991, {Third Reference Catalogue of
  Bright Galaxies. Volume I: Explanations and references. Volume II: Data for
  galaxies between 0$^{h}$ and 12$^{h}$. Volume III: Data for galaxies between
  12$^{h}$ and 24$^{h}$.}

\bibitem[{{Dekel} \& {Woo}(2003)}]{dekelwoo2003}
{Dekel}, A., \& {Woo}, J. 2003, \mnras, 344, 1131

\bibitem[{{Dressler} {et~al.}(1987){Dressler}, {Lynden-Bell}, {Burstein},
  {Davies}, {Faber}, {Terlevich}, \& {Wegner}}]{dressleretal1987}
{Dressler}, A., {Lynden-Bell}, D., {Burstein}, D., {Davies}, R.~L., {Faber},
  S.~M., {Terlevich}, R., \& {Wegner}, G. 1987, \apj, 313, 42

\bibitem[{{Dutton} {et~al.}(2012){Dutton}, {Mendel}, \&
  {Simard}}]{duttonetal2012}
{Dutton}, A.~A., {Mendel}, J.~T., \& {Simard}, L. 2012, \mnras, 422, L33

\bibitem[{{Emsellem} {et~al.}(2004)}]{emsellemetal2004}
{Emsellem}, E., {et~al.} 2004, \mnras, 352, 721

\bibitem[{{Erb} {et~al.}(2006){Erb}, {Shapley}, {Pettini}, {Steidel}, {Reddy},
  \& {Adelberger}}]{erbetal2006}
{Erb}, D.~K., {Shapley}, A.~E., {Pettini}, M., {Steidel}, C.~C., {Reddy},
  N.~A., \& {Adelberger}, K.~L. 2006, \apj, 644, 813

\bibitem[{{Faber} {et~al.}(1977){Faber}, {Burstein}, \&
  {Dressler}}]{faberetal1977}
{Faber}, S.~M., {Burstein}, D., \& {Dressler}, A. 1977, \aj, 82, 941

\bibitem[{{Faber} {et~al.}(2007)}]{faberetal2007}
{Faber}, S.~M., {et~al.} 2007, \apj, 665, 265

\bibitem[{{Ferr{\'e}-Mateu} {et~al.}(2012){Ferr{\'e}-Mateu}, {Vazdekis},
  {Trujillo}, {S{\'a}nchez-Bl{\'a}zquez}, {Ricciardelli}, \& {de la
  Rosa}}]{ferremateuetal2012}
{Ferr{\'e}-Mateu}, A., {Vazdekis}, A., {Trujillo}, I.,
  {S{\'a}nchez-Bl{\'a}zquez}, P., {Ricciardelli}, E., \& {de la Rosa}, I.~G.
  2012, \mnras, 423, 632

\bibitem[{{Finkelstein} {et~al.}(2011)}]{finkelsteinetal2011}
{Finkelstein}, S.~L., {et~al.} 2011, \apj, 729, 140

\bibitem[{{Fisher} {et~al.}(1995){Fisher}, {Franx}, \&
  {Illingworth}}]{fisheretal1995}
{Fisher}, D., {Franx}, M., \& {Illingworth}, G. 1995, \apj, 448, 119

\bibitem[{{Franx} \& {Illingworth}(1990)}]{franxillingworth1990}
{Franx}, M., \& {Illingworth}, G. 1990, \apjl, 359, L41

\bibitem[{{Gallazzi} {et~al.}(2005){Gallazzi}, {Charlot}, {Brinchmann},
  {White}, \& {Tremonti}}]{gallazzietal2005}
{Gallazzi}, A., {Charlot}, S., {Brinchmann}, J., {White}, S.~D.~M., \&
  {Tremonti}, C.~A. 2005, \mnras, 362, 41

\bibitem[{{Garnett} {et~al.}(1999){Garnett}, {Shields}, {Peimbert},
  {Torres-Peimbert}, {Skillman}, {Dufour}, {Terlevich}, \&
  {Terlevich}}]{garnettetal1999}
{Garnett}, D.~R., {Shields}, G.~A., {Peimbert}, M., {Torres-Peimbert}, S.,
  {Skillman}, E.~D., {Dufour}, R.~J., {Terlevich}, E., \& {Terlevich}, R.~J.
  1999, \apj, 513, 168

\bibitem[{{Gorgas} {et~al.}(1990){Gorgas}, {Efstathiou}, \& {Aragon
  Salamanca}}]{gorgasetal1990}
{Gorgas}, J., {Efstathiou}, G., \& {Aragon Salamanca}, A. 1990, \mnras, 245,
  217

\bibitem[{{Graves} {et~al.}(2009){Graves}, {Faber}, \&
  {Schiavon}}]{gravesetal2009}
{Graves}, G.~J., {Faber}, S.~M., \& {Schiavon}, R.~P. 2009, \apj, 693, 486

\bibitem[{{Graves} {et~al.}(2010){Graves}, {Faber}, \&
  {Schiavon}}]{gravesetal2010}
---. 2010, \apj, 721, 278

\bibitem[{{Graves} {et~al.}(2007){Graves}, {Faber}, {Schiavon}, \&
  {Yan}}]{gravesetal2007}
{Graves}, G.~J., {Faber}, S.~M., {Schiavon}, R.~P., \& {Yan}, R. 2007, \apj,
  671, 243

\bibitem[{{Graves} \& {Schiavon}(2008)}]{gravesschiavon2008}
{Graves}, G.~J., \& {Schiavon}, R.~P. 2008, \apjs, 177, 446

\bibitem[{{Greene} {et~al.}(2012){Greene}, {Murphy}, {Comerford}, {Gebhardt},
  \& {Adams}}]{greeneetal2012}
{Greene}, J.~E., {Murphy}, J.~D., {Comerford}, J.~M., {Gebhardt}, K., \&
  {Adams}, J.~J. 2012, \apj, 750, 32

\bibitem[{{Groenewegen}(1999)}]{groenewegen1999}
{Groenewegen}, M.~A.~T. 1999, in IAU Symposium, Vol. 191, Asymptotic Giant
  Branch Stars, ed. T.~{Le Bertre}, A.~{Lebre}, \& C.~{Waelkens}, 535

\bibitem[{{Harris} {et~al.}(1999){Harris}, {Harris}, \&
  {Poole}}]{harrisetal1999}
{Harris}, G.~L.~H., {Harris}, W.~E., \& {Poole}, G.~B. 1999, \aj, 117, 855

\bibitem[{{Harris} {et~al.}(2007){Harris}, {Harris}, {Layden}, \&
  {Wehner}}]{harrisetal2007}
{Harris}, W.~E., {Harris}, G.~L.~H., {Layden}, A.~C., \& {Wehner}, E.~M.~H.
  2007, \apj, 666, 903

\bibitem[{{Henry} {et~al.}(2000){Henry}, {Edmunds}, \&
  {K{\"o}ppen}}]{henryetal2000}
{Henry}, R.~B.~C., {Edmunds}, M.~G., \& {K{\"o}ppen}, J. 2000, \apj, 541, 660

\bibitem[{{Hill} {et~al.}(2008)}]{hilletal2008a}
{Hill}, G.~J., {et~al.} 2008, in Society of Photo-Optical Instrumentation
  Engineers (SPIE) Conference Series, Vol. 7014, Society of Photo-Optical
  Instrumentation Engineers (SPIE) Conference Series

\bibitem[{{Hilz} {et~al.}(2013){Hilz}, {Naab}, \& {Ostriker}}]{hilzetal2013}
{Hilz}, M., {Naab}, T., \& {Ostriker}, J.~P. 2013, \mnras, 429, 2924

\bibitem[{{Hilz} {et~al.}(2012){Hilz}, {Naab}, {Ostriker}, {Thomas}, {Burkert},
  \& {Jesseit}}]{hilzetal2012}
{Hilz}, M., {Naab}, T., {Ostriker}, J.~P., {Thomas}, J., {Burkert}, A., \&
  {Jesseit}, R. 2012, \mnras, 425, 3119

\bibitem[{{Huang} {et~al.}(2012){Huang}, {Ho}, {Peng}, {Li}, \&
  {Barth}}]{huangetal2013a}
{Huang}, S., {Ho}, L.~C., {Peng}, C.~Y., {Li}, Z.-Y., \& {Barth}, A.~J. 2012,
  \apj, accepted

\bibitem[{{Huang} {et~al.}(2013){Huang}, {Ho}, {Peng}, {Li}, \&
  {Barth}}]{huangetal2013b}
---. 2013, \apjl, 768, L28

\bibitem[{{Izotov} {et~al.}(1999){Izotov}, {Chaffee}, {Foltz}, {Green},
  {Guseva}, \& {Thuan}}]{izotovthuan1999}
{Izotov}, Y.~I., {Chaffee}, F.~H., {Foltz}, C.~B., {Green}, R.~F., {Guseva},
  N.~G., \& {Thuan}, T.~X. 1999, \apj, 527, 757

\bibitem[{{Jablonka} {et~al.}(2007){Jablonka}, {Gorgas}, \&
  {Goudfrooij}}]{jablonkaetal2007}
{Jablonka}, P., {Gorgas}, J., \& {Goudfrooij}, P. 2007, \aap, 474, 763

\bibitem[{{Jiang} {et~al.}(2012){Jiang}, {van Dokkum}, {Bezanson}, \&
  {Franx}}]{jiangetal2012}
{Jiang}, F., {van Dokkum}, P., {Bezanson}, R., \& {Franx}, M. 2012, \apjl, 749,
  L10

\bibitem[{{Johansson} {et~al.}(2012){Johansson}, {Thomas}, \&
  {Maraston}}]{johanssonetal2012}
{Johansson}, J., {Thomas}, D., \& {Maraston}, C. 2012, \mnras, 421, 1908

\bibitem[{{Kalirai} {et~al.}(2006)}]{kaliraietal2006}
{Kalirai}, J.~S., {et~al.} 2006, \apj, 648, 389

\bibitem[{{Kelson}(2003)}]{kelson2003}
{Kelson}, D.~D. 2003, \pasp, 115, 688

\bibitem[{{Kelson} {et~al.}(2006){Kelson}, {Illingworth}, {Franx}, \& {van
  Dokkum}}]{kelsonetal2006}
{Kelson}, D.~D., {Illingworth}, G.~D., {Franx}, M., \& {van Dokkum}, P.~G.
  2006, \apj, 653, 159

\bibitem[{{Kobayashi}(2004)}]{kobayashi2004}
{Kobayashi}, C. 2004, \mnras, 347, 740

\bibitem[{{Kobayashi} \& {Arimoto}(1999)}]{kobayashiarimoto1999}
{Kobayashi}, C., \& {Arimoto}, N. 1999, \apj, 527, 573

\bibitem[{{Kormendy} {et~al.}(2009){Kormendy}, {Fisher}, {Cornell}, \&
  {Bender}}]{kormendyetal2009}
{Kormendy}, J., {Fisher}, D.~B., {Cornell}, M.~E., \& {Bender}, R. 2009, \apjs,
  182, 216

\bibitem[{{Korn} {et~al.}(2005){Korn}, {Maraston}, \& {Thomas}}]{kornetal2005}
{Korn}, A.~J., {Maraston}, C., \& {Thomas}, D. 2005, \aap, 438, 685

\bibitem[{{Kuntschner} {et~al.}(2010)}]{kuntschneretal2010}
{Kuntschner}, H., {et~al.} 2010, \mnras, 408, 97

\bibitem[{{Lackner} \& {Gunn}(2012)}]{lacknergunn2012}
{Lackner}, C.~N., \& {Gunn}, J.~E. 2012, \mnras, 421, 2277

\bibitem[{{Larson}(1974)}]{larson1974}
{Larson}, R.~B. 1974, \mnras, 169, 229

\bibitem[{{L{\"a}sker} {et~al.}(2013){L{\"a}sker}, {van den Bosch}, {van de
  Ven}, {Ferreras}, {La Barbera}, {Vazdekis}, \&
  {Falc{\'o}n-Barroso}}]{laskeretal2013}
{L{\"a}sker}, R., {van den Bosch}, R.~C.~E., {van de Ven}, G., {Ferreras}, I.,
  {La Barbera}, F., {Vazdekis}, A., \& {Falc{\'o}n-Barroso}, J. 2013, \mnras,
  in press (arXiv:1305.5542)

\bibitem[{{Maeder}(1992)}]{maeder1992}
{Maeder}, A. 1992, \aap, 264, 105

\bibitem[{{Mandelbaum} {et~al.}(2005)}]{mandelbaumetal2005}
{Mandelbaum}, R., {et~al.} 2005, \mnras, 361, 1287

\bibitem[{{Mannucci} {et~al.}(2010){Mannucci}, {Cresci}, {Maiolino}, {Marconi},
  \& {Gnerucci}}]{mannuccietal2010}
{Mannucci}, F., {Cresci}, G., {Maiolino}, R., {Marconi}, A., \& {Gnerucci}, A.
  2010, \mnras, 408, 2115

\bibitem[{{Marigo} {et~al.}(1998){Marigo}, {Bressan}, \&
  {Chiosi}}]{marigoetal1998}
{Marigo}, P., {Bressan}, A., \& {Chiosi}, C. 1998, \aap, 331, 564

\bibitem[{{Matteucci}(1986)}]{matteucci1986}
{Matteucci}, F. 1986, \mnras, 221, 911

\bibitem[{{Matteucci} \& {Greggio}(1986)}]{matteuccigreggio1986}
{Matteucci}, F., \& {Greggio}, L. 1986, \aap, 154, 279

\bibitem[{{Mehlert} {et~al.}(2003){Mehlert}, {Thomas}, {Saglia}, {Bender}, \&
  {Wegner}}]{mehlertetal2003}
{Mehlert}, D., {Thomas}, D., {Saglia}, R.~P., {Bender}, R., \& {Wegner}, G.
  2003, \aap, 407, 423

\bibitem[{{Meynet} \& {Maeder}(2002)}]{meynetmaeder2002}
{Meynet}, G., \& {Maeder}, A. 2002, \aap, 390, 561

\bibitem[{{Mouhcine} \& {Lan{\c c}on}(2003)}]{mouhcinelancon2003}
{Mouhcine}, M., \& {Lan{\c c}on}, A. 2003, \mnras, 338, 572

\bibitem[{{Murphy} {et~al.}(2011){Murphy}, {Gebhardt}, \&
  {Adams}}]{murphyetal2011}
{Murphy}, J.~D., {Gebhardt}, K., \& {Adams}, J.~J. 2011, \apj, 729, 129

\bibitem[{{Naab} {et~al.}(2009){Naab}, {Johansson}, \&
  {Ostriker}}]{naabetal2009}
{Naab}, T., {Johansson}, P.~H., \& {Ostriker}, J.~P. 2009, \apjl, 699, L178

\bibitem[{{Naab} \& {Ostriker}(2009)}]{naabostriker2009}
{Naab}, T., \& {Ostriker}, J.~P. 2009, \apj, 690, 1452

\bibitem[{{Newman} {et~al.}(2012){Newman}, {Ellis}, {Bundy}, \&
  {Treu}}]{newmanetal2012}
{Newman}, A.~B., {Ellis}, R.~S., {Bundy}, K., \& {Treu}, T. 2012, \apj, 746,
  162

\bibitem[{{Ogando} {et~al.}(2005){Ogando}, {Maia}, {Chiappini}, {Pellegrini},
  {Schiavon}, \& {da Costa}}]{ogandoetal2005}
{Ogando}, R.~L.~C., {Maia}, M.~A.~G., {Chiappini}, C., {Pellegrini}, P.~S.,
  {Schiavon}, R.~P., \& {da Costa}, L.~N. 2005, \apjl, 632, L61

\bibitem[{{Oser} {et~al.}(2012){Oser}, {Naab}, {Ostriker}, \&
  {Johansson}}]{oseretal2012}
{Oser}, L., {Naab}, T., {Ostriker}, J.~P., \& {Johansson}, P.~H. 2012, \apj,
  744, 63

\bibitem[{{Oser} {et~al.}(2010){Oser}, {Ostriker}, {Naab}, {Johansson}, \&
  {Burkert}}]{oseretal2010}
{Oser}, L., {Ostriker}, J.~P., {Naab}, T., {Johansson}, P.~H., \& {Burkert}, A.
  2010, \apj, 725, 2312

\bibitem[{{Pipino} {et~al.}(2010){Pipino}, {D'Ercole}, {Chiappini}, \&
  {Matteucci}}]{pipinoetal2010}
{Pipino}, A., {D'Ercole}, A., {Chiappini}, C., \& {Matteucci}, F. 2010, \mnras,
  407, 1347

\bibitem[{{Price} {et~al.}(2011){Price}, {Phillipps}, {Huxor}, {Smith}, \&
  {Lucey}}]{priceetal2011}
{Price}, J., {Phillipps}, S., {Huxor}, A., {Smith}, R.~J., \& {Lucey}, J.~R.
  2011, \mnras, 411, 2558

\bibitem[{{Pu} \& {Han}(2011)}]{puhan2011}
{Pu}, S.-B., \& {Han}, Z.-W. 2011, Research in Astronomy and Astrophysics, 11,
  909

\bibitem[{{Pu} {et~al.}(2010){Pu}, {Saglia}, {Fabricius}, {Thomas}, {Bender},
  \& {Han}}]{puetal2010}
{Pu}, S.~B., {Saglia}, R.~P., {Fabricius}, M.~H., {Thomas}, J., {Bender}, R.,
  \& {Han}, Z. 2010, \aap, 516, A4

\bibitem[{{Rawle} {et~al.}(2008){Rawle}, {Smith}, {Lucey}, \&
  {Swinbank}}]{rawleetal2008}
{Rawle}, T.~D., {Smith}, R.~J., {Lucey}, J.~R., \& {Swinbank}, A.~M. 2008,
  \mnras, 389, 1891

\bibitem[{{Rejkuba} {et~al.}(2005){Rejkuba}, {Greggio}, {Harris}, {Harris}, \&
  {Peng}}]{rejkubaetal2005}
{Rejkuba}, M., {Greggio}, L., {Harris}, W.~E., {Harris}, G.~L.~H., \& {Peng},
  E.~W. 2005, \apj, 631, 262

\bibitem[{{Renzini} \& {Voli}(1981)}]{renzinivoli1981}
{Renzini}, A., \& {Voli}, M. 1981, \aap, 94, 175

\bibitem[{{Salasnich} {et~al.}(2000){Salasnich}, {Girardi}, {Weiss}, \&
  {Chiosi}}]{salasnichetal2000}
{Salasnich}, B., {Girardi}, L., {Weiss}, A., \& {Chiosi}, C. 2000, \aap, 361,
  1023

\bibitem[{{S{\'a}nchez-Bl{\'a}zquez} {et~al.}(2007){S{\'a}nchez-Bl{\'a}zquez},
  {Forbes}, {Strader}, {Brodie}, \& {Proctor}}]{sanchez-blazquezetal2007}
{S{\'a}nchez-Bl{\'a}zquez}, P., {Forbes}, D.~A., {Strader}, J., {Brodie}, J.,
  \& {Proctor}, R. 2007, \mnras, 377, 759

\bibitem[{{S{\'a}nchez-Bl{\'a}zquez} {et~al.}(2006){S{\'a}nchez-Bl{\'a}zquez},
  {Gorgas}, {Cardiel}, \& {Gonz{\'a}lez}}]{sanchezblazquezetal2006}
{S{\'a}nchez-Bl{\'a}zquez}, P., {Gorgas}, J., {Cardiel}, N., \& {Gonz{\'a}lez},
  J.~J. 2006, \aap, 457, 787

\bibitem[{{Sarzi} {et~al.}(2006)}]{sarzietal2006}
{Sarzi}, M., {et~al.} 2006, \mnras, 366, 1151

\bibitem[{{Sarzi} {et~al.}(2010)}]{sarzietal2010}
---. 2010, \mnras, 402, 2187

\bibitem[{{Schiavon}(2007)}]{schiavon2007}
{Schiavon}, R.~P. 2007, \apjs, 171, 146

\bibitem[{{Scott} {et~al.}(2009)}]{scottetal2009}
{Scott}, N., {et~al.} 2009, \mnras, 398, 1835

\bibitem[{{Scott} {et~al.}(2012)}]{scottetal2013}
---. 2012, ArXiv e-prints

\bibitem[{{Smith} {et~al.}(2009){Smith}, {Lucey}, {Hudson}, \&
  {Bridges}}]{smithetal2009}
{Smith}, R.~J., {Lucey}, J.~R., {Hudson}, M.~J., \& {Bridges}, T.~J. 2009,
  \mnras, 398, 119

\bibitem[{{Spinrad} \& {Taylor}(1971)}]{spinradtaylor1971}
{Spinrad}, H., \& {Taylor}, B.~J. 1971, \apjs, 22, 445

\bibitem[{{Spolaor} {et~al.}(2010){Spolaor}, {Kobayashi}, {Forbes}, {Couch}, \&
  {Hau}}]{spolaoretal2010}
{Spolaor}, M., {Kobayashi}, C., {Forbes}, D.~A., {Couch}, W.~J., \& {Hau},
  G.~K.~T. 2010, \mnras, 408, 272

\bibitem[{{Strader} {et~al.}(2013)}]{straderetal2013}
{Strader}, J., {et~al.} 2013, ArXiv e-prints

\bibitem[{{Strateva} {et~al.}(2001)}]{stratevaetal2001}
{Strateva}, I., {et~al.} 2001, \aj, 122, 1861

\bibitem[{{Szomoru} {et~al.}(2012){Szomoru}, {Franx}, \& {van
  Dokkum}}]{szomoruetal2012}
{Szomoru}, D., {Franx}, M., \& {van Dokkum}, P.~G. 2012, \apj, 749, 121

\bibitem[{{Thomas} {et~al.}(2005){Thomas}, {Maraston}, {Bender}, \& {Mendes de
  Oliveira}}]{thomasetal2005}
{Thomas}, D., {Maraston}, C., {Bender}, R., \& {Mendes de Oliveira}, C. 2005,
  \apj, 621, 673

\bibitem[{{Thomas} {et~al.}(2011)}]{thomasetal2011}
{Thomas}, J., {et~al.} 2011, \mnras, 415, 545

\bibitem[{{Toft} {et~al.}(2007)}]{toftetal2007}
{Toft}, S., {et~al.} 2007, \apj, 671, 285

\bibitem[{{Trager} {et~al.}(2000{\natexlab{a}}){Trager}, {Faber}, {Worthey}, \&
  {Gonz{\'a}lez}}]{trageretal2000b}
{Trager}, S.~C., {Faber}, S.~M., {Worthey}, G., \& {Gonz{\'a}lez}, J.~J.
  2000{\natexlab{a}}, \aj, 120, 165

\bibitem[{{Trager} {et~al.}(2000{\natexlab{b}}){Trager}, {Faber}, {Worthey}, \&
  {Gonz{\'a}lez}}]{trageretal2000a}
---. 2000{\natexlab{b}}, \aj, 119, 1645

\bibitem[{{Trager} \& {Somerville}(2009)}]{tragersomerville2009}
{Trager}, S.~C., \& {Somerville}, R.~S. 2009, \mnras, 395, 608

\bibitem[{{Tremonti} {et~al.}(2004)}]{tremontietal2004}
{Tremonti}, C.~A., {et~al.} 2004, \apj, 613, 898

\bibitem[{{Tripicco} \& {Bell}(1995)}]{tripiccobell1995}
{Tripicco}, M.~J., \& {Bell}, R.~A. 1995, \aj, 110, 3035

\bibitem[{{Trujillo} {et~al.}(2009){Trujillo}, {Cenarro}, {de
  Lorenzo-C{\'a}ceres}, {Vazdekis}, {de la Rosa}, \& {Cava}}]{trujilloetal2009}
{Trujillo}, I., {Cenarro}, A.~J., {de Lorenzo-C{\'a}ceres}, A., {Vazdekis}, A.,
  {de la Rosa}, I.~G., \& {Cava}, A. 2009, \apjl, 692, L118

\bibitem[{{Trujillo} {et~al.}(2006)}]{trujilloetal2006}
{Trujillo}, I., {et~al.} 2006, \mnras, 373, L36

\bibitem[{{Valentinuzzi} {et~al.}(2010)}]{valentinuzzietal2010}
{Valentinuzzi}, T., {et~al.} 2010, \apj, 712, 226

\bibitem[{{van de Sande} {et~al.}(2011)}]{vandesandeetal2011}
{van de Sande}, J., {et~al.} 2011, \apjl, 736, L9

\bibitem[{{van den Bosch} {et~al.}(2012){van den Bosch}, {Gebhardt},
  {G{\"u}ltekin}, {van de Ven}, {van der Wel}, \&
  {Walsh}}]{vandenboschetal2012}
{van den Bosch}, R.~C.~E., {Gebhardt}, K., {G{\"u}ltekin}, K., {van de Ven},
  G., {van der Wel}, A., \& {Walsh}, J.~L. 2012, \nat, 491, 729

\bibitem[{{van der Wel} {et~al.}(2008){van der Wel}, {Holden}, {Zirm}, {Franx},
  {Rettura}, {Illingworth}, \& {Ford}}]{vanderweletal2008}
{van der Wel}, A., {Holden}, B.~P., {Zirm}, A.~W., {Franx}, M., {Rettura}, A.,
  {Illingworth}, G.~D., \& {Ford}, H.~C. 2008, \apj, 688, 48

\bibitem[{{van Dokkum} {et~al.}(2008)}]{vandokkumetal2008}
{van Dokkum}, P.~G., {et~al.} 2008, \apjl, 677, L5

\bibitem[{{van Dokkum} {et~al.}(2010)}]{vandokkumetal2010}
---. 2010, \apj, 709, 1018

\bibitem[{{Vazdekis} {et~al.}(2010){Vazdekis}, {S{\'a}nchez-Bl{\'a}zquez},
  {Falc{\'o}n-Barroso}, {Cenarro}, {Beasley}, {Cardiel}, {Gorgas}, \&
  {Peletier}}]{vazdekisetal2010}
{Vazdekis}, A., {S{\'a}nchez-Bl{\'a}zquez}, P., {Falc{\'o}n-Barroso}, J.,
  {Cenarro}, A.~J., {Beasley}, M.~A., {Cardiel}, N., {Gorgas}, J., \&
  {Peletier}, R.~F. 2010, \mnras, 404, 1639

\bibitem[{{Venn} {et~al.}(2004){Venn}, {Irwin}, {Shetrone}, {Tout}, {Hill}, \&
  {Tolstoy}}]{vennetal2004}
{Venn}, K.~A., {Irwin}, M., {Shetrone}, M.~D., {Tout}, C.~A., {Hill}, V., \&
  {Tolstoy}, E. 2004, \aj, 128, 1177

\bibitem[{{Weijmans} {et~al.}(2009)}]{weijmansetal2009}
{Weijmans}, A.-M., {et~al.} 2009, \mnras, 398, 561

\bibitem[{{Wetzel} {et~al.}(2012){Wetzel}, {Tinker}, \&
  {Conroy}}]{wetzeletal2012}
{Wetzel}, A.~R., {Tinker}, J.~L., \& {Conroy}, C. 2012, \mnras, 424, 232

\bibitem[{{Williams} {et~al.}(2010){Williams}, {Quadri}, {Franx}, {van Dokkum},
  {Toft}, {Kriek}, \& {Labb{\'e}}}]{williamsetal2010}
{Williams}, R.~J., {Quadri}, R.~F., {Franx}, M., {van Dokkum}, P., {Toft}, S.,
  {Kriek}, M., \& {Labb{\'e}}, I. 2010, \apj, 713, 738

\bibitem[{{Worthey}(2004)}]{worthey2004}
{Worthey}, G. 2004, \aj, 128, 2826

\bibitem[{{Worthey} {et~al.}(1992){Worthey}, {Faber}, \&
  {Gonzalez}}]{wortheyetal1992}
{Worthey}, G., {Faber}, S.~M., \& {Gonzalez}, J.~J. 1992, \apj, 398, 69

\bibitem[{{Worthey} {et~al.}(1994){Worthey}, {Faber}, {Gonzalez}, \&
  {Burstein}}]{wortheyetal1994}
{Worthey}, G., {Faber}, S.~M., {Gonzalez}, J.~J., \& {Burstein}, D. 1994,
  \apjs, 94, 687

\bibitem[{{Worthey} {et~al.}(2013){Worthey}, {Tang}, \&
  {Serven}}]{wortheyetal2013}
{Worthey}, G., {Tang}, B., \& {Serven}, J. 2013, \apj, submitted
  (arXiv:1306.2603)

\bibitem[{{Yan}(2011)}]{yan2011}
{Yan}, R. 2011, \aj, 142, 153

\bibitem[{{Yan} \& {Blanton}(2012)}]{yanblanton2012}
{Yan}, R., \& {Blanton}, M.~R. 2012, \apj, 747, 61

\bibitem[{{Yang} {et~al.}(2007){Yang}, {Mo}, {van den Bosch}, {Pasquali}, {Li},
  \& {Barden}}]{yangetal2007}
{Yang}, X., {Mo}, H.~J., {van den Bosch}, F.~C., {Pasquali}, A., {Li}, C., \&
  {Barden}, M. 2007, \apj, 671, 153

\bibitem[{{Yates} {et~al.}(2013){Yates}, {Henriques}, {Thomas}, {Kauffmann},
  {Johansson}, \& {White}}]{yatesetal2013}
{Yates}, R.~M., {Henriques}, B., {Thomas}, P.~A., {Kauffmann}, G., {Johansson},
  J., \& {White}, S.~D.~M. 2013, \mnras, submitted (arXiv:1305.7231)

\bibitem[{{Yoachim} {et~al.}(2010){Yoachim}, {Ro{\v s}kar}, \&
  {Debattista}}]{yoachimetal2010}
{Yoachim}, P., {Ro{\v s}kar}, R., \& {Debattista}, V.~P. 2010, \apjl, 716, L4

\bibitem[{{York} {et~al.}(2000)}]{yorketal2000}
{York}, D.~G., {et~al.} 2000, \aj, 120, 1579

\bibitem[{{Zhu} {et~al.}(2010){Zhu}, {Blanton}, \& {Moustakas}}]{zhuetal2010}
{Zhu}, G., {Blanton}, M.~R., \& {Moustakas}, J. 2010, \apj, 722, 491

\end{thebibliography}

\end{document}